\documentclass[11pt,letterpaper]{article}
\date{}
\usepackage{setspace}
\usepackage[english]{babel}
\usepackage[nottoc]{tocbibind}
\usepackage{amsmath,amssymb}
\usepackage{amsthm}
\usepackage{forest}
\usepackage[authoryear,round]{natbib}
\usepackage{algorithm}
\usepackage{algpseudocode}
\usepackage{booktabs}
\usepackage{xcolor}
\usepackage{threeparttable}
\usepackage{physics}
\usepackage[small,compact]{titlesec} 
\usepackage{multirow}
\usepackage{xcolor}

\usetikzlibrary{fit,positioning}

\usepackage[right=1in,left=1in,top=1in,bottom=1in]{geometry}

\usepackage{subcaption}
\usepackage{mathptmx}
\usepackage{amsmath}
\usepackage{graphicx}
\usepackage[colorlinks=true, allcolors=blue]{hyperref}
\usepackage{pifont}
\usepackage[toc,page]{appendix}

\newtheorem{proposition}{Proposition}
\newtheorem{remark}{Remark}

\newtheorem{assumption}{Assumption}
\newtheorem{definition}{Definition}
\usepackage{amsmath}
\usepackage{amssymb}
\usepackage{amsthm}

\newcommand{\squishlist}{
   \begin{list}{$\bullet$}
    { \setlength{\itemsep}{0pt} \setlength{\parsep}{1pt}
      \setlength{\topsep}{1pt} \setlength{\partopsep}{1pt}
      \setlength{\leftmargin}{1.5em} \setlength{\labelwidth}{1em}
      \setlength{\labelsep}{0.5em} } }

\newcommand{\squishlisttwo}{
   \begin{list}{$\bullet$}
    { \setlength{\itemsep}{0pt} \setlength{\parsep}{0pt}
      \setlength{\topsep}{0pt} \setlength{\partopsep}{0pt}
      \setlength{\leftmargin}{1em} \setlength{\labelwidth}{1.5em}
      \setlength{\labelsep}{0.5em} } }

\newcommand{\squishend}{
    \end{list}  }

\begin{document}
%\kern-2.5cm
\title{The Privacy–Utility Trade-Off of Location Tracking in Ad Personalization}

\author{
      Mohammad Mosaffa\thanks{We would like to thank the participants of the 2025 PhD student workshop at Cornell University and the 2026 Artificial Intelligence in Management (AIM) Conference for their feedback. We also thank Emaad Manzoor for the detailed comments, which have significantly improved the paper. Please address all correspondence to: mm3322@cornell.edu and or83@cornell.edu.}\\
      Cornell University \\
      \and 
      Omid Rafieian\footnotemark[1]\\
        Cornell University\\ 
 }

\date{}
\maketitle

\begin{abstract}
Firms collect vast amounts of behavioral and geographical data on individuals. While behavioral data captures an individual’s digital footprint, geographical data reflects their physical footprint. Given the significant privacy risks associated with combining these data sources, it is crucial to understand their respective value and whether they act as complements or substitutes in achieving firms’ business objectives. In this paper, we combine economic theory, machine learning, and causal inference to quantify the value of geographical data, the extent to which behavioral data can substitute it, and the mechanisms through which it benefits firms. Using data from a leading in-app advertising platform in a large Asian country, we document that geographical data is most valuable in the early “cold-start” stage, when behavioral histories are limited. In this stage, geographical data \emph{complements} behavioral data, improving targeting performance by almost 20\%. As users accumulate richer behavioral histories, however, the role of geographical data shifts: it becomes largely \emph{substitutable}, as behavioral data alone captures the relevant heterogeneity. These results highlight a central privacy–utility trade-off in ad personalization and inform managerial decisions about when location tracking creates value.

\noindent{\bf Keywords:} Privacy, Personalization, Advertising, Informational Complementarity, Policy Evaluation, Spatial Statistics

\end{abstract}

\thispagestyle{empty}

\newpage

\section{Introduction}
Mobile devices dominate the digital economy, reshaping how consumers interact with media, commerce, and advertising. Nearly all Americans own a mobile phone, and nine in ten use a smartphone \citep{pew2024smartphone}, which creates an ideal environment for targeted marketing. In response, mobile advertising now accounts for more than two-thirds of all digital ad spending in the United States, surpassing \$200 billion in 2024 \citep{emarketer2024mobile}. Much of this growth stems from advances in tracking technologies that allow advertisers to monitor users’ digital and physical activities and tailor messages to individual consumers. While these capabilities have made mobile advertising a central channel for marketers, they have also heightened privacy concerns, raising questions about whether the economic value of such data practices justifies the associated privacy risks.

Privacy concerns arise most directly from two types of data collected by advertising platforms: \emph{behavioral data}, such as click histories or in-app purchases, and \emph{geographical data}, such as GPS coordinates or IP-based locations \citep{ghose2019mobile}. Behavioral data captures a user’s digital footprint within apps, while geographical data captures their physical footprint—their movements and proximity in the real world. Geographical information is particularly sensitive, as even a few location points can be sufficient to re-identify individuals \citep{de2013unique}. When combined with behavioral data, these risks amplify: the two data types complement each other in enabling more precise inference of personal attributes and increasing the likelihood of re-identification, even when datasets are anonymized \citep{de2018privacy}.

Given the complementarity of geographical and behavioral information in amplifying privacy risks, it is crucial to examine whether a similar complementarity exists in the factor that motivates firms to continue collecting them—the economic value they create. In particular, it remains unclear whether behavioral and geographical data act as \textit{substitutes} or \textit{complements} in generating economic value for advertising platforms. Addressing this question requires first establishing how to measure the value of each data type, then comparing their relative contributions and interactions, and finally identifying the mechanisms through which geographical data may provide distinct informational value. These considerations motivate the following research questions:

\begin{enumerate}
\item How can we measure the value of geographical data in ad personalization relative to behavioral data?  
\item To what extent do geographical data generate value once behavioral data are available, and how does this value change as behavioral data accumulate?  
\item Through which mechanisms do geographical data provide distinct information?
\end{enumerate}

We face several challenges in answering these questions. The first challenge is conceptual: how should we define the value of information? A common approach focuses on prediction accuracy, for example, whether adding geographical data improves click-through forecasts. However, predictive gains alone do not guarantee better advertising outcomes \citep{ascarza2018retention}. What ultimately matters is whether available information improves decision quality. Accordingly, we draw on economic theory to define the value of geographical and behavioral data through their impact on decision outcomes. This framework allows us to assess whether the two sources act as \emph{complements}, where their joint use delivers gains beyond the sum of their individual effects, or as \emph{substitutes}, where the contribution of one diminishes once the other is available. For example, if behavioral data indicate that a user is a New York Knicks fan, knowing that the user is in New York City is valuable for advertising tickets to a home game, so the two information sources may act as complements. However, for online Knicks merchandise sold online and shipped nationwide, location adds little value once behavioral interest is known, so the two sources may act as substitutes. This distinction is central, as collecting information beyond its marginal value imposes privacy costs without improving outcomes, whereas complementary information can justify broader data use.

%The second challenge is empirical. Evaluating decision value requires prescribing policies conditional on each information set, which in turn demands models that reflect the structure of the data. While Behavioral information is inherently temporal, capturing evolving histories of impressions, geographical information is high-dimensional and spatially correlated across regions. Ignoring these features risks misspecified policies and biased conclusions. To address this challenge, we employ machine learning architectures designed to capture both temporal and spatial structure, including long short-term memory (LSTM) networks with attention mechanisms.

The second challenge is empirical. Our objective is to assess whether geographical data provides incremental decision value beyond behavioral data. This requires a behavioral model that fully captures the dynamic structure of user histories; otherwise, estimated gains from geographical data may reflect behavioral misspecification rather than informational value. Because behavioral data are inherently temporal, we leverage a best-in-class Long Short-Term Memory (LSTM) network with an attention mechanism to capture temporal dependencies and heterogeneous relevance across past interactions. 

%The third challenge is methodological. Policy evaluation requires counterfactual outcomes that are not directly observed: we only see user responses under the policies actually deployed, not under alternative targeting strategies. To overcome this limitation, we apply inverse propensity scoring (IPS) \citep{horvitz1952generalization}, which recovers counterfactual values by reweighting observed outcomes using assignment probabilities. A key feature of our setting is the platform’s proportional auction mechanism, which allocates ads in proportion to the quality-adjusted bid. This mechanism generates plausibly exogenous variation, supporting reliable propensity score estimation and credible counterfactual evaluation.

The third challenge is statistical. Our objective is to test whether two sources of information act as substitutes or complements, which requires comparing policy values under alternative information sets. This, in turn, demands counterfactual outcomes that are not directly observed: we only observe user responses under the policies actually deployed, not under alternative targeting strategies. To address this limitation, we apply inverse propensity scoring (IPS) \citep{horvitz1952generalization}, which recovers counterfactual policy values by reweighting observed outcomes using assignment probabilities. A key feature of our setting is the platform’s quasi-proportional auction mechanism, which is directly observed in the data and allocates ads in proportion to the quality-adjusted bid. This mechanism provides plausibly exogenous variation and supports reliable propensity score estimation and credible statistical inference.

Together, these challenges motivate a unified framework that combines economic theory to define information value through policy comparisons, machine learning to model user outcomes from complex behavioral and geographical data, and causal inference to enable counterfactual evaluation via IPS. We apply this framework to data from a leading mobile advertising network in a major Asian country, covering 10.3 million impressions from 439{,}157 users over a 10-day period. We consider four targeting scenarios: a benchmark \(X^{\varnothing}\) that uses only contextual information (e.g., app category or time of day); \(X^G\), which augments context with geographical data such as exact latitude and longitude; \(X^B\) augments contextual features with behavioral information from users’ prior impression and click histories; and \(X^{GB}\), which combines both geographical and behavioral information. Differences in data structure across scenarios motivate the use of \emph{best-in-class} models, with gradient-boosted trees (XGBoost) for scenarios based on static and cross-sectional features (\(X^{\varnothing}\) and \(X^G\)) and LSTM networks with attention for scenarios that incorporate sequential behavioral histories (\(X^B\) and \(X^{GB}\)). Leveraging the observed quasi-proportional auction mechanism to support credible propensity score estimation, we apply IPS to recover counterfactual click-through rates (CTRs) and formally quantify the value of behavioral and geographical information, providing a statistical test of whether the two sources act as complements or substitutes.

At the aggregate level, we compare all targeting regimes to the baseline CTR. Targeting without user-level data (\(X^{\varnothing}\)) increases CTR by 20.44\%. Adding geographical information (\(X^G\)) raises CTR by 28.7\%, behavioral information (\(X^B\)) by 30.3\%, and combining both (\(X^{GB}\)) yields an improvement of 41.5\%. Although the joint regime achieves the highest performance, aggregate results alone do not provide systematic evidence that geographical and behavioral information act as complements or substitutes. This motivates examining heterogeneity as behavioral information accumulates.

As users receive more ad impressions, advertisers observe richer behavioral histories, which change both the value of behavioral information and its interaction with geographical information. In the earliest stage (1--2 impressions), behavioral information is minimal, and performance is driven largely by geographical information, with the combined regime performing similarly to geographical data alone. In the intermediate stage (approximately 5--25 impressions), behavioral information accumulates but remains sparse. In this range, geographical and behavioral information act as complements, and using both information sources yields gains beyond either source alone. Once users exceed 25 impressions, behavioral information becomes sufficiently rich to capture preferences independently, and geographical information becomes a substitute, adding little incremental value. Overall, these results show that geographical information evolves with impression depth: complementary when it is sparse, and substitutable once it is rich.

Our findings have direct implications for managers and policymakers. Geographical information creates value primarily in the short run, acting as a temporary complement when behavioral information is sparse, but becomes substitutable as user histories accumulate. For firms, this implies that geo-targeting can be useful during the cold-start phase but delivers little incremental benefit once behavioral models are established. For regulators, the results underscore a key trade-off: location information entails substantial privacy risks while offering limited long-term value in digital advertising. Firms should therefore reassess the strategic role of geographical information, using it selectively when behavioral histories are unavailable and reducing reliance on it as richer user profiles emerge.

In summary, our paper makes several contributions to the literature. First, while prior research has emphasized the value of either behavioral or geographical data in isolation, we provide a systematic comparison of the two. Second, we develop a unified framework that combines economic theory to define information value, machine learning to model advertising responses, and causal inference to evaluate counterfactual outcomes. This framework allows managers and policymakers to assess whether different information sources act as complements or substitutes, providing a practical tool to evaluate targeting decisions while accounting for privacy costs. Substantively, we show that both geographical and behavioral data create value, but their roles evolve as behavioral information accumulates: geographical data complements behavioral data when histories are sparse and becomes substitutable once behavioral information is rich. This pattern implies that the marginal value of geo-targeting is highest when behavioral information is sparse and declines as behavioral histories become richer, raising questions about the privacy–utility trade-off.

The remainder of the paper proceeds as follows. We review related work on the value of information, advertising and privacy, and spatial economics in \S\ref{sec:literature}. We introduce the institutional setting of mobile advertising and describe the large-scale dataset in \S\ref{sec:data}. We then formalize the problem within a decision-under-uncertainty framework to define the value of information in \S\ref{sec:ProblemDef}. Next, we first highlight key challenges and then present our empirical framework in \S\ref{sec:strategy}, where we define complementarity and substitutability and combine machine-learning–based click prediction with inverse propensity scoring for counterfactual evaluation. We report empirical findings on the value of behavioral and geographical data and their heterogeneity in \S\ref{sec:Results}, and we examine the mechanisms behind these patterns using a residualized spatial autocorrelation test in \S\ref{sec:geo_mechanismm}. We discuss implications for data use, privacy, and targeting efficiency in \S\ref{sec:Managerial}, and we conclude by summarizing our contributions and outlining directions for future research in \S\ref{sec:conclusion}. %\mohammad{Added}

%These findings indicate that geo-targeting offers short-term benefits but limited long-term usefulness, highlighting the need to weigh its contribution against potential privacy concerns.

%The remainder of the paper is organized as follows. \S\ref{sec:literature} reviews related work on the value of information, advertising and privacy, and spatial economics. \S\ref{sec:data} introduces the institutional setting of mobile advertising and describes our large-scale dataset. \S\ref{sec:ProblemDef} formalizes the problem within a decision-under-uncertainty framework to define the value of information and highlight key challenges, while \S\ref{sec:strategy} presents our identification strategy, beginning with the definitions of complementarity and substitutability and leveraging machine learning for click prediction and inverse propensity scoring for counterfactual evaluation. \S\ref{sec:Results} reports empirical findings on the value of behavioral and geographical data and their heterogeneity. \S\ref{sec:Managerial} discusses implications for data use, privacy, and targeting efficiency, and \S\ref{sec:conclusion} concludes with a summary of contributions and directions for future research.

\section{Related Work}\label{sec:literature}

Our work relates to a broad literature on how the value of information is defined and measured for decision-making under uncertainty. Theoretically, \citet{blackwell1953comparison} provides the benchmark by ordering signals according to whether they increase expected reward across all decision problems (the Blackwell order), and \citet{borgers2013signals} extend this framework to multiple signals by characterizing when information sources act as complements or substitutes based on whether one signal’s marginal value rises or falls in the presence of another.  Empirically, \citet{rossi1996value} quantify the value of alternative information sets for direct marketing by comparing profits from targeted couponing based on purchase histories and demographic characteristics. Complementing this perspective, \citet{kim2022selecting} show that the value extracted from rich data depends on the chosen level of data granularity and model specification, formalizing the bias–variance trade-off in information use. Subsequent work examines how data access and targeting policies affect outcomes using causal strategies. Using randomized experiments, \citet{ascarza2018retention} compares targeting rules based on churn risk versus treatment-effect lift, \citet{cui2019learning} manipulate the availability of information shown to consumers, and \citet{wernerfelt2025estimating} experimentally restrict access to offsite behavioral data to estimate its incremental value. Other studies exploit natural experiments, such as \citet{aridor2024evaluating}, who use Apple’s App Tracking Transparency as an exogenous shock to behavioral data access to identify its impact on advertising outcomes. Finally, off-policy evaluation methods recover policy values from logged or observational data, as in \citet{rafieian2023ai} and \citet{rafieian2023optimizing}. We extend this literature by proposing a unified framework that systematically compares multiple information sources and empirically tests whether they act as complements or substitutes through their marginal contributions to decision outcomes in high-dimensional settings. %\omid{There are a few issues here. Ascarza and Wernerfelt can be separated as citations. Also, there are more papers focusing on the value of different pieces of information. I'm sure Dennis Zhang and Eric Bradlow have papers on targeting and the use of different pieces of information.} \mohammad{Revised!}

Second, our study connects to the literature on privacy and personalized advertising, particularly in user tracking and engagement modeling. In display advertising, early work by \citet{goldfarb2011online} shows that ad intrusiveness and privacy sensitivity significantly affect engagement, with well-targeted yet subtle ads performing best. Subsequent research demonstrates that personalized ads can increase relevance but also intensify privacy concerns: \citet{tucker2014social} show that personalization in social networks raises demand for stronger privacy controls, while \citet{acquisti2016economics} formalize the trade-offs firms face between personalization benefits and consumer resistance to data collection. More recent studies examine the consequences of restricting data access, with \citet{johnson2020consumer} quantifying the revenue losses from consumer opt-outs and \citet{Rafieian2021} showing that privacy restrictions reduce targeting efficiency and may affect market competition. In this regard, despite well-documented privacy risks of location data \citep{de2013unique}, relatively few studies directly examine the value of geographical information. Closely related to our work, \citet{narang2025privacy} quantify the predictive value of geo-tracking data for forecasting consumer visits. We extend this line of research by moving beyond predictive accuracy and developing a unified framework for substitutability/complementarity between geographical and behavioral data that equips us to more precisely assess the privacy--utility trade-off. %\omid{We should definitely cite Unnati's paper here. We can say this is a closely related paper to ours and we extend it by going beyond predictive accuracy improvement of geo data and establishing a framework for substitutability/complementarity that equips us to better assess privacy-utility trade-off.} \mohammad{Revised}

%In this regard, despite well-documented privacy risks of location data \citep{de2013unique}, the privacy–utility trade-off of geographical data remains unexamined. We fill this gap by empirically assessing when geographical data add value and when they do not justify their privacy costs. 

Third, a line of empirical research in marketing and economics studies geographic dependence in outcomes using methods from spatial econometrics. Early contributions introduce diagnostics for spatial autocorrelation. \citet{moran1950notes} proposes Moran’s \( I \) as a global measure of spatial dependence in outcomes, while \citet{geary1954contiguity} develops Geary’s \( C \) to capture more localized spatial variation. Subsequent work applies these diagnostics to substantive economic settings. \citet{bronnenberg2001unobserved} model spatial dependence in market shares and promotions across neighboring markets to account for correlated, unobserved retailer actions. Focusing on longer-run demand patterns, \citet{bronnenberg2009brand} document persistent geographic structure in brand demand, showing that brands retain higher shares near their historical origins. Related research examines diffusion and contagion by exploiting geographic or network adjacency. A body of empirical research in marketing studies how spatial structure shapes consumer behavior and diffusion outcomes \citep{bradlow2005spatial}. At finer levels of granularity, \citet{larson2005exploratory} show that spatial layout affects consumer exposure and choice through physical shopping paths. This perspective extends to diffusion and contagion, where researchers exploit geographic or network adjacency: \citet{manchanda2008role} use physician networks to separate targeted communication from peer influence, \citet{iyengar2011opinion} study social networks to identify opinion leadership effects, and \citet{bollinger2012peer} use zip-code–level exposure to quantify neighborhood spillovers in solar adoption. Despite growing evidence of substantial spatial heterogeneity in advertising effectiveness, as documented by \citet{luo2025mapping}, existing work largely focuses on detecting and modeling spatial correlation in outcomes rather than evaluating geographical data as an input to ad personalization. We address this gap by assessing the value of geographical information for personalization and by proposing a residual spatial autocorrelation (RSA) framework that decomposes spatial correlation conditional on behavioral data to pin down the channel through which geographical information creates value. %\omid{There is a JMR paper by Luo and Ranjan (2025). The distinction could be stronger by saying that we assess the value of geo data in ad personalization and propose RSA that enables us to pin down the channel through which geo information creates value by decomposing the spatial correlation. } \mohammad{Revised}

\section{Setting \& Data}\label{sec:data}

We start by describing the institutional setting of the mobile advertising platform (\S\ref{ssec:setting}). Next, we detail the dataset, including impressions, clicks, and contextual variables (\S\ref{ssec:data}). We then explain our sampling strategy and report summary statistics (\S\ref{ssec:summary}). Finally, we describe the train–test split design used to evaluate model performance (\S\ref{ssec:split}).

\subsection{Setting}\label{ssec:setting}
Our data are sourced from a leading mobile in-app advertising platform in a large Asian country, which held over 85\% of the mobile advertising market during the time of our study. The platform serves as an intermediary between advertisers and mobile apps (publishers) and is responsible for delivering over 50 million ad impressions daily. This marketplace consists of four primary players:
\begin{itemize}
  \item \textbf{Users} are mobile app consumers who generate impressions and may choose to click on displayed ads.
  \item \textbf{Publishers} are app developers that integrate ads into their apps and monetize based on ad clicks.
  \item \textbf{Advertisers} design banner ads and specify per-click bids. They can target users based on variables such as province, smartphone brand, app category, internet service provider (ISP), hour of the day, and connectivity type. The ad network does not support detailed personalized targeting.
  \item \textbf{Platform or ad network} manages real-time auctions to match impressions with ads. Ads are placed as bottom banners and are refreshed every minute. Only clicks result in payment under a cost-per-click (CPC) scheme.
\end{itemize}

\begin{figure}[htp!]
    \centering
    \includegraphics[width=0.75\linewidth]{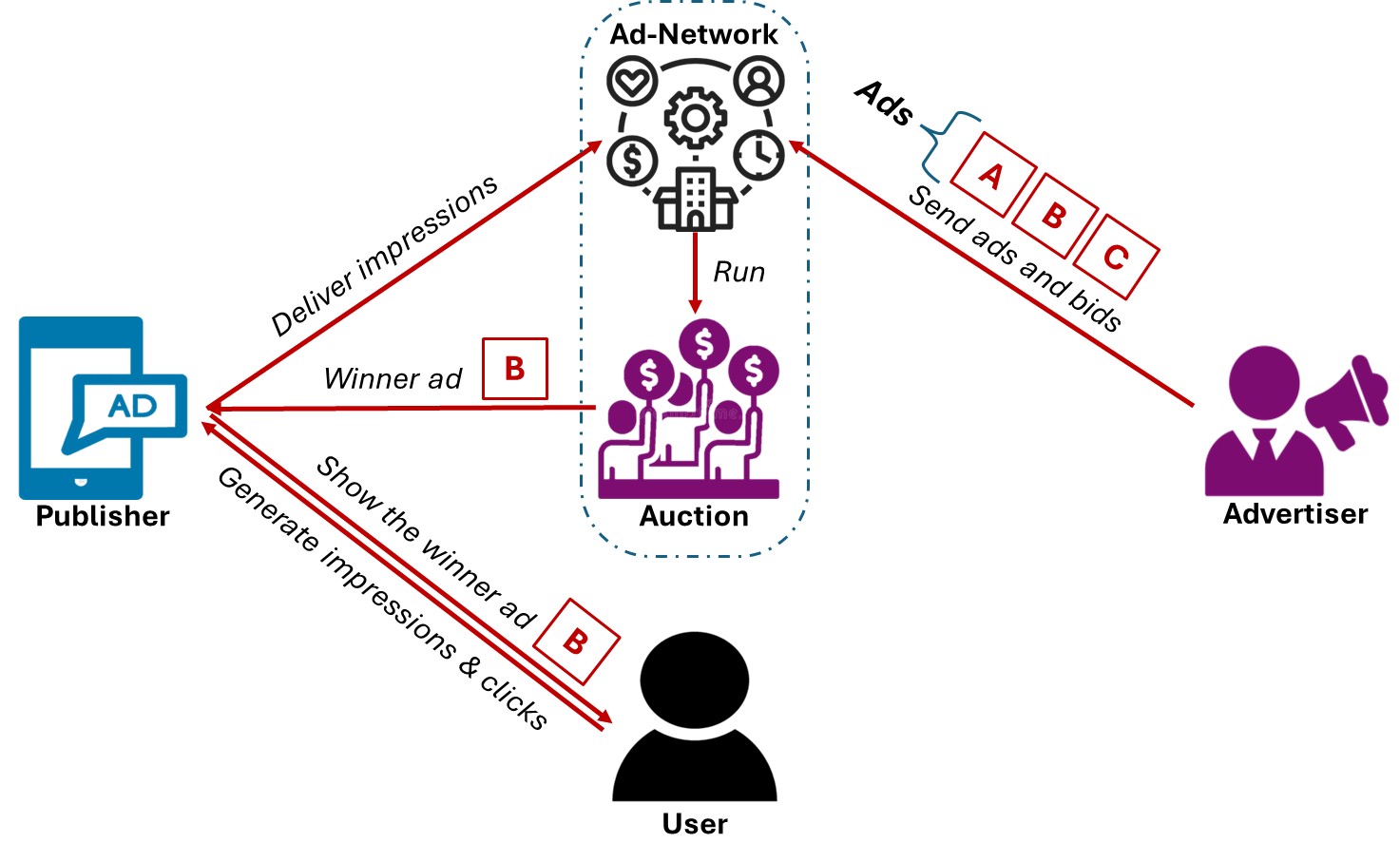}
    \caption{Schematic of the mobile in-app advertising marketplace.}
    \label{fig:market_schema}
\end{figure}

Figure~\ref{fig:market_schema} illustrates the structure of the in-app advertising marketplace. When a user opens an app, the platform initiates a real-time auction among eligible ads, allocating impressions through a quasi-proportional rule \citep{mirrokni2010quasi}: ads with higher bid–quality products are assigned higher probabilities of being shown, but selection remains stochastic. For example, if two ads have bid–quality products of 2 and 1, the first ad is twice as likely to be shown as the second, but both ads retain strictly positive probabilities of exposure. As a result, propensity scores are non-zero by design, which ensures overlap across ads. Importantly, quality scores are fixed and not personalized at the user level, so allocation probabilities do not depend on unobserved user characteristics, implying unconfounded assignment by design. If the user remains active beyond one minute, a new impression is generated, and a new auction is run.

\subsection{Data}\label{ssec:data}

Following the setting explained above, we have data on all impressions and click outcomes observed over a 30-day period from September 30, 2015, to October 30, 2015. During this period, we observe a total of 1,594,831,699 ad impressions and 14,373,293 clicks, resulting in an overall click-through rate (CTR) of approximately 0.90\%. For each impression, the dataset includes detailed information across several dimensions.

For each impression, the dataset records a comprehensive set of variables. First,
we observe \textit{(1) Timestamp}, capturing the exact time of the impression, and \textit{(2) AAID}, the Android Advertising ID, a user-resettable unique device identifier that enables anonymous tracking across applications. At the ad-delivery level, we record \textit{(3) AppID}, the identifier of the app displaying the ad, and \textit{(4) CreativeID}, the identifier of the specific ad creative shown. We also observe \textit{(5) Bid}, the advertiser’s submitted bid amount (fixed throughout the sample period), and \textit{(6) CPC}, the cost-per-click charged in the event of a click. Geographical attributes include \textit{(7) Latitude}, \textit{(8) Longitude}, and \textit{(9) Province}, which are available for most impressions. Further contextual information comprises \textit{(10) Connectivity}, indicating whether the user was on Wi-Fi or cellular data, \textit{(11) Brand}, the smartphone manufacturer, \textit{(12) MSP}, the mobile service provider, and \textit{(13) ISP}, the internet service provider. Finally, \textit{(14) Click} is a binary indicator equal to one if the user clicked on the ad and zero otherwise.

A key aspect of this dataset is that it is sourced directly from the ad platform and includes the complete set of variables that advertisers could potentially use for targeting. As such, we observe the same information available to both the platform and advertisers at the time of impression delivery. This mitigates common concerns
in observational studies around hidden targeting mechanisms or unobserved confounding. In our context, the completeness of the data supports modeling assumptions such as conditional ignorability, which are crucial for later analyses involving counterfactual analysis and policy evaluation. %Further details on the data are provided in Appendix \S\ref{sec:detailData}.

\subsection{Sampling and Summary Statistics}\label{ssec:summary}
To enable user-level analysis while maintaining computational tractability, we restrict attention to impressions from the top 10 ads served between October~20 and October~30, yielding an initial sample of 16{,}662{,}783 impressions. We address missing values, particularly in variables that serve as key features in our analysis, by excluding records without geographic information, specifically latitude and longitude. After this filtering, the working dataset contains 10{,}336{,}703 impressions and has no remaining missing values in any other columns. From this dataset, we identify new users who joined the platform during the first three days of the window, October 20 to October 22, and follow their subsequent activity through October 30. This design allows us to observe user responsiveness from the start of their platform interaction, yielding a cohort of 439,157 unique users.

We now present some summary statistics for key categorical variables in the dataset. Table~\ref{tab:categorical_summary} reports, for each variable, the number of unique categories, the share of impressions associated with the top three values, and the number of non-missing observations.

\begin{table}[ht]
\centering
\caption{Summary Statistics for the Categorical Variables}
\label{tab:categorical_summary}
\begin{tabular}{l r r r r r}
\toprule
\textbf{Variable} & \textbf{Number of categories} & \multicolumn{3}{c}{\textbf{Share of top categories}} & \textbf{Number of impressions} \\
\cmidrule(lr){3-5}
 &  & \textbf{1st} & \textbf{2nd} & \textbf{3rd} &  \\
\midrule
App &     9,515 & 26.14\% &  9.46\% &  4.54\% & 10,336,703 \\
Ad      &        10 & 22.33\% & 13.28\% & 12.67\% & 10,336,703 \\
Unique User        &   439,157 & 0.02\% & 0.02\% & 0.02\% & 10,336,703 \\
Smartphone brand           &         7 & 41.94\% & 30.59\% &  9.21\% & 10,336,703 \\
Connectivity Type                &         2 & 54.65\% & 45.35\% &  0.00\% & 10,336,703 \\
ISP             &         8 & 61.09\% & 26.93\% &  5.24\% & 10,336,703 \\
Province            &       828 & 11.21\% &  8.78\% &  7.33\% & 10,336,703 \\
\bottomrule
\end{tabular}
\end{table}

We observe a total of 9,515 unique apps, with the top three accounting for a sizable share of total impressions. Among the ten ads included by design, exposure is uneven, with the most frequently shown ad representing over 22\% of impressions. The distribution of user identifiers, based on Android Advertising IDs, is highly diffuse, with no single user accounting for more than 0.02\% of total impressions. Other variables, such as smartphone brand, connectivity type, ISP, and province, exhibit varying degrees of concentration, reflecting both common patterns and localized variation in usage across the population. 

While Table~\ref{tab:categorical_summary} highlights variation in exposure across contextual features, it does not reveal how user responsiveness may differ across behavioral or geographical information. To explore this further, we present descriptive evidence on how CTR varies with user history and spatial location. These patterns help motivate the relevance of behavioral and geographical data for downstream modeling tasks.

\subsubsection{Behavioral Heterogeneity}
We present two descriptive results that illustrate how behavioral patterns shape user CTR. First, we examine how the length of a user’s exposure history relates to CTR. Second, we analyze the relationship between past click behavior and the likelihood of future clicks.

\begin{figure}[htbp]
\centering
\begin{subfigure}[t]{0.455\textwidth}
    \centering
    \includegraphics[width=\linewidth]{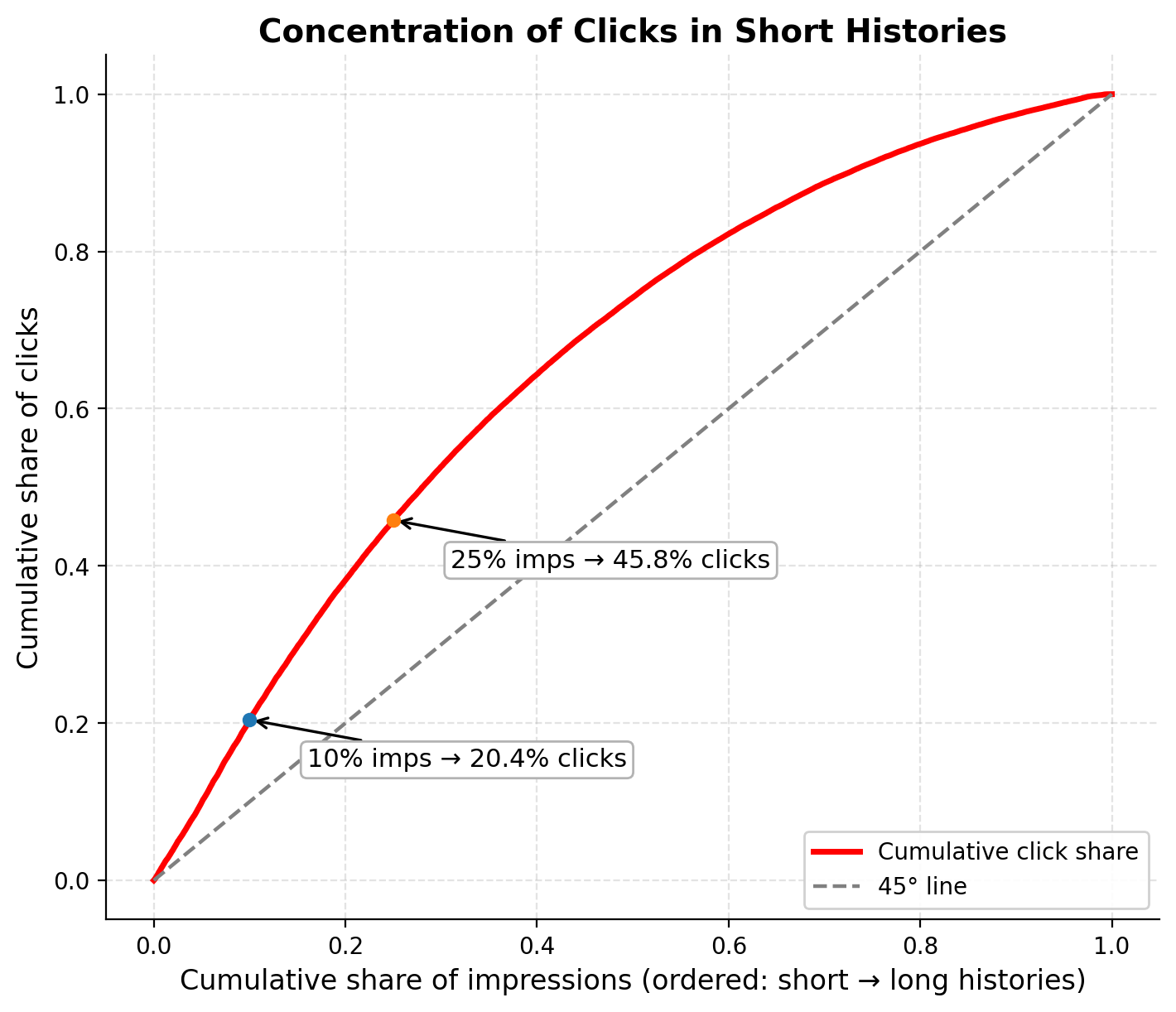}
    \caption{Concentration of clicks in short histories}
    \label{fig:Fig_Lorenz_History}
\end{subfigure}%
\hfill
\begin{subfigure}[t]{0.545\textwidth}
    \centering
    \includegraphics[width=\linewidth]{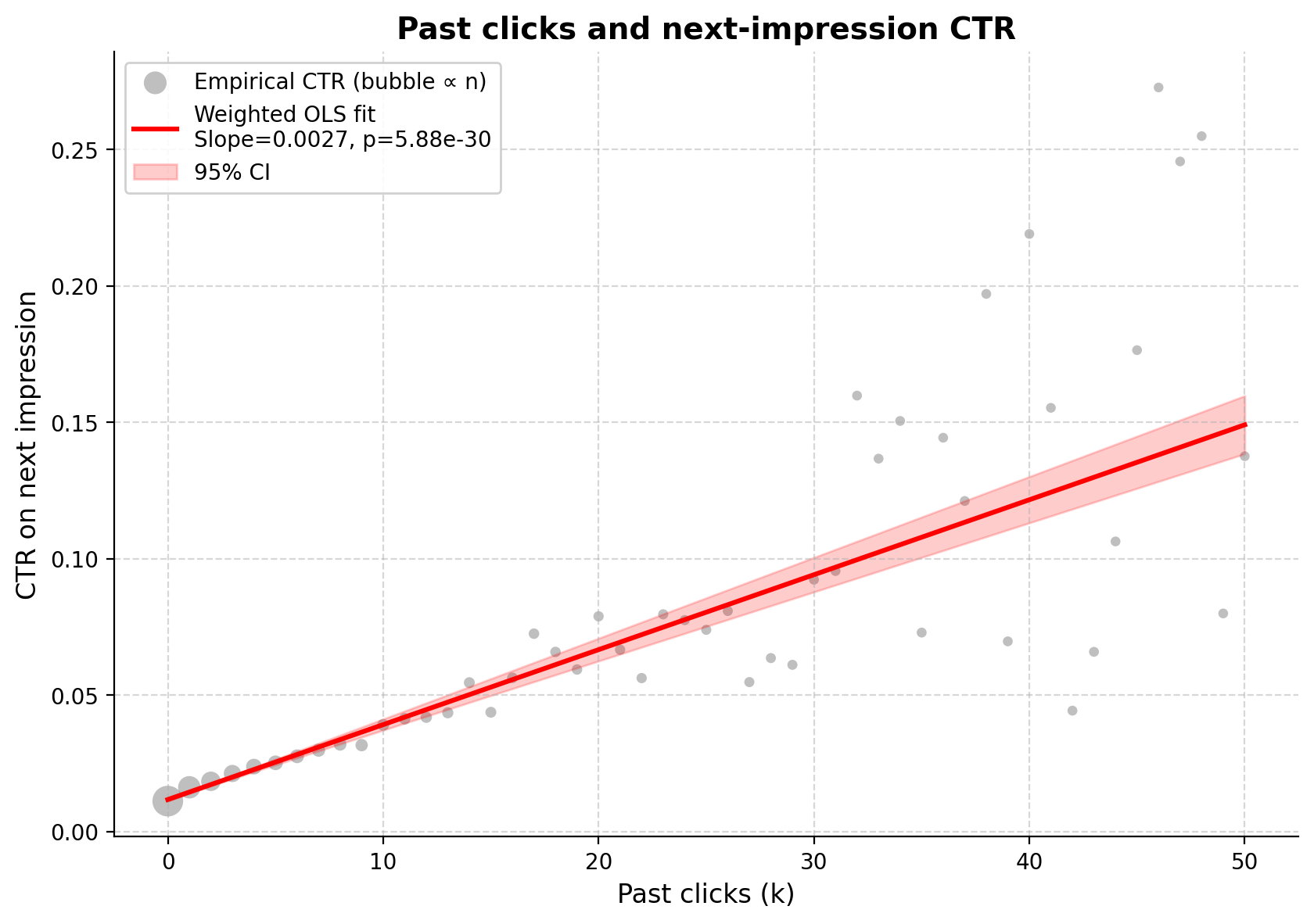}
    \caption{Persistence in click behavior}
    \label{fig:Fig_Past_Click}
\end{subfigure}
\vspace{0.5em}
\caption{Behavioral heterogeneity: (a) Clicks are concentrated in early exposure histories; (b) users with prior clicks are more likely to click again.}
\label{fig:heterogeneity_plots}
\end{figure}

Figure~\ref{fig:Fig_Lorenz_History} plots the cumulative share of impressions against the cumulative share of clicks, where users are ordered by the number of prior impressions they have seen, so that impressions are ranked from early to late in a user’s exposure history, and each point on the curve compares the share of total impressions up to that history length with the share of total clicks they generate. The curve lies well above the $45^\circ$ line: a disproportionate share of clicks comes from early exposures (e.g., the first 25\% of impressions, corresponding to short histories, generate nearly half of all clicks). Figure~\ref{fig:Fig_Past_Click} shows the probability of a click on the next impression, $\Pr(\text{click}_{t+1}=1 \mid \text{prior clicks}=k)$, as a function of the number of past clicks $k$. The weighted linear fit has a positive and statistically significant slope: users who have accumulated more clicks are more likely to click again on the next impression.\footnote{Final impressions (with no $t{+}1$) are excluded. These patterns are descriptive and do not imply causal effects.}

Taken together, these descriptive results reveal two dimensions of behavioral heterogeneity. First, Figure~\ref{fig:Fig_Lorenz_History} shows that CTR declines as exposure histories lengthen, indicating the need to account for user trajectories. Second, Figure~\ref{fig:Fig_Past_Click} shows persistence in click behavior: users with more prior clicks are more likely to click again, reflecting heterogeneity in underlying propensities and highlighting the predictive value of behavioral history.

\subsubsection{Geographical Heterogeneity }
\begin{figure}[htp!]
    \centering
    \includegraphics[width=0.6\linewidth]{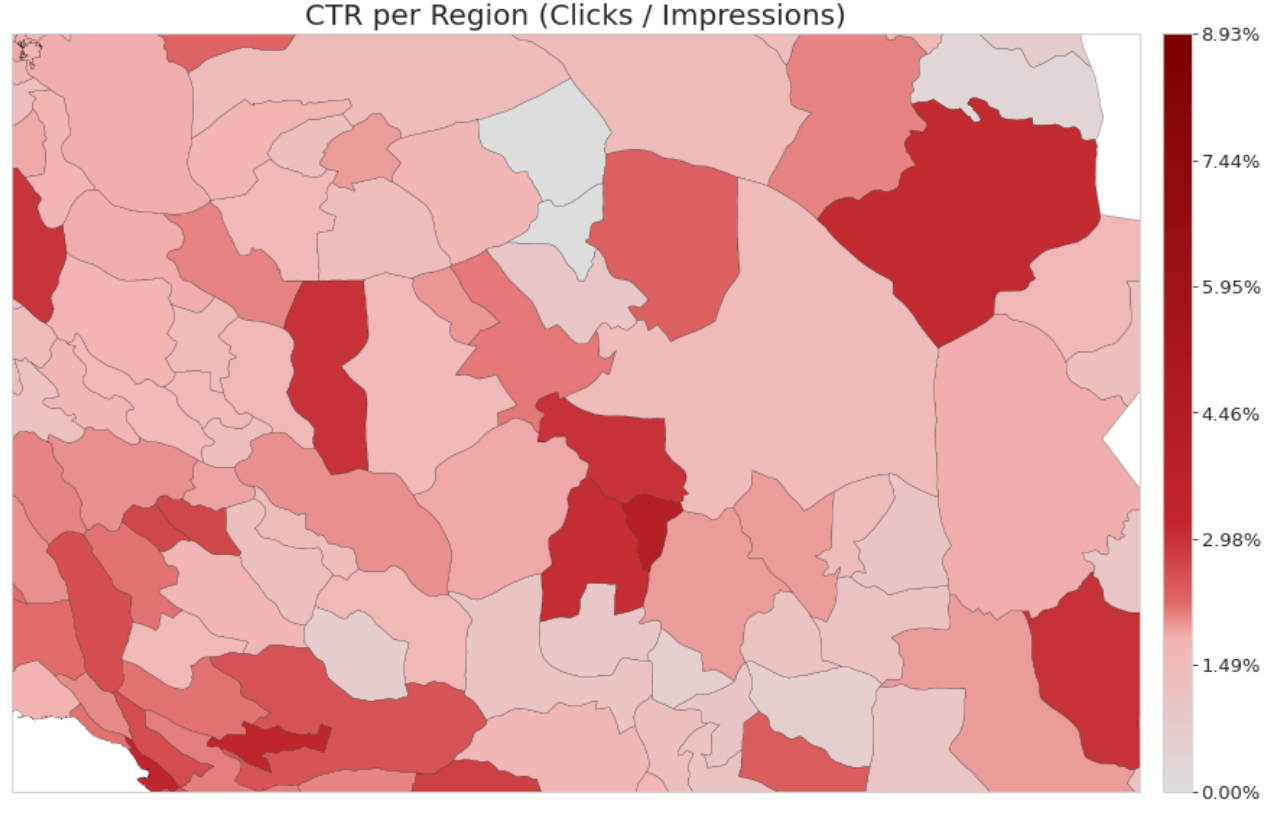}
    \caption{Average CTR by County}
    \label{fig:ctr_county}
\end{figure}

We now turn to geographical patterns of responsiveness to examine whether user responses exhibit systematic spatial dependence. To do so, we group impressions by county and compute the average CTR within each spatial unit. Specifically, for a given county \( c \), we calculate \( \text{CTR}_c = \frac{1}{N_c} \sum_{i \in c} y_i, \) where \( y_i \in \{0, 1\} \) indicates whether impression \( i \) resulted in a click, and \( N_c \) is the number of impressions observed in county \( c \). Figure~\ref{fig:ctr_county} visualizes the resulting CTR values as a choropleth map.

The figure reveals pronounced spatial correlation in responsiveness: counties with high (low) average CTRs tend to be geographically proximate to other counties with similarly high (low) CTRs, which generate visible regional clusters rather than isolated pockets of responsiveness. This spatial structure indicates that user engagement is not independently distributed across space but instead varies smoothly across neighboring locations. Such correlation suggests that geographic context captures shared latent factors, such as local demographics or socioeconomic context, that influence users' likelihood of clicking on ads.

%The figure shows that responsiveness varies meaningfully across counties. Some regions exhibit higher average CTRs, while others show lower responsiveness, even after aggregating over large numbers of impressions. These differences may reflect underlying geographic heterogeneity\omid{I think the point here should be more about spatial correlation rather than spatial variation/heterogeneity}, shaped by socioeconomic context, that influences users' likelihood of clicking on ads.

\subsection{Train-Test Split Strategy}\label{ssec:split}

Finally, we split the data into training and test sets to evaluate out-of-sample predictive performance and policy effectiveness. To avoid information leakage, the split is performed at the user level rather than the impression level: each user’s full history is contained entirely within either the training or the test partition, allowing us to assess generalization to previously unseen users.

\begin{figure}[ht]
    \centering
    \includegraphics[width=0.75\linewidth]{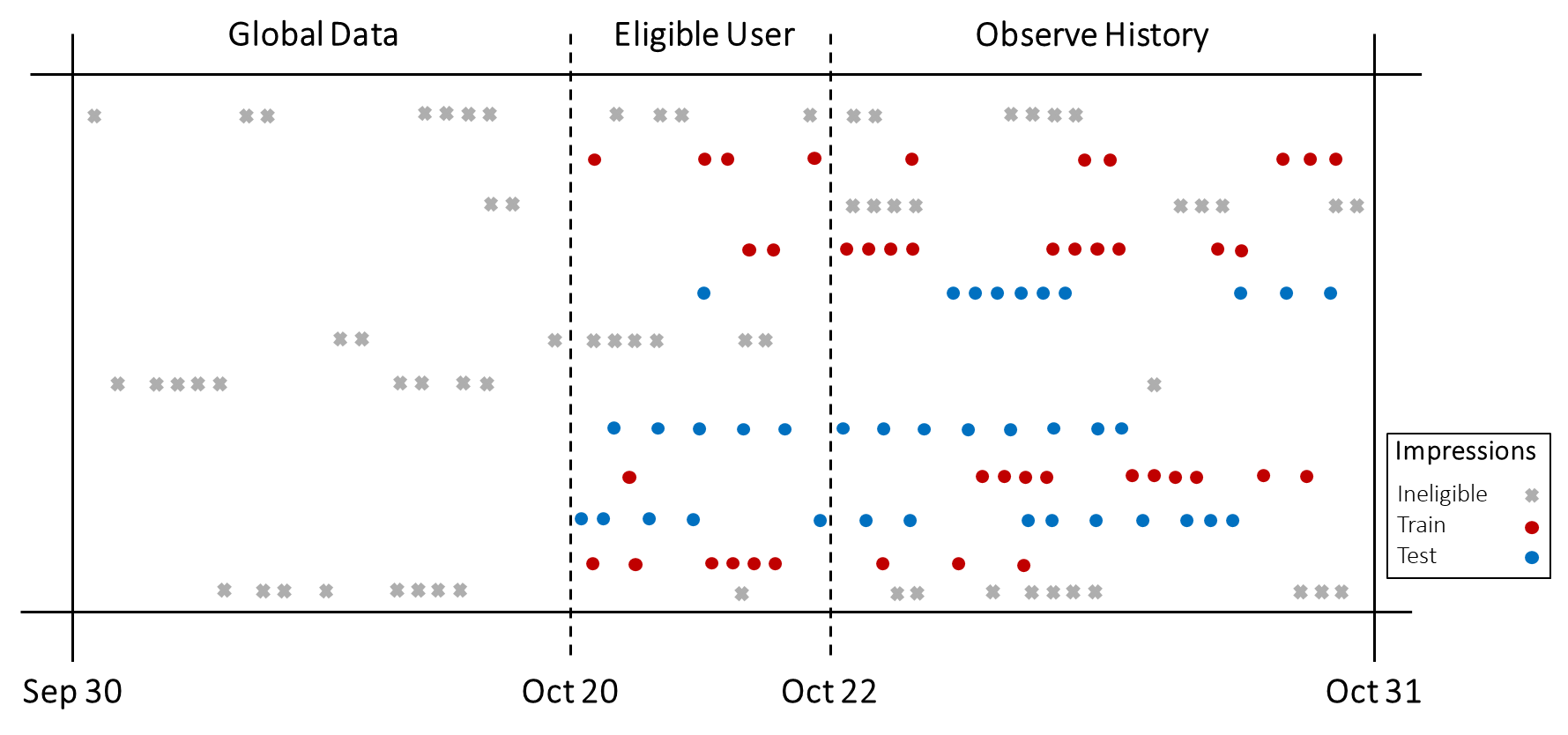}
    \caption{Schema for Data Generation and User-Based Train-Test Split}
    \label{fig:data_split}
\end{figure}

Figure~\ref{fig:data_split} illustrates the procedure. A user becomes eligible once they first appear in the platform’s logs, after which we track their impression history. Each user is then randomly assigned to either the training set (60\%, 263,494 users; 6,202,022 impressions) or the test set (40\%, 175,663 users; 4,134,681 impressions). Gray dots mark ineligible impressions that precede a user’s first appearance, while red and blue dots denote impressions assigned to the training and test sets, respectively. The resulting split maintains similar distributions of activity and click behavior across both partitions.

\section{Problem Definition}\label{sec:ProblemDef} 
Advertising platforms operate in environments where advertising decisions must be made under uncertainty about user preferences. Consider the case of an ad network deciding which ad to show to a particular user. The central goal of the platform is to maximize a reward measure, such as advertising revenue or user engagement. At the same time, the platform does not fully observe the user's type but observes a signal through observed features. As a result, the platform faces a decision-making problem under uncertainty, where the challenge lies in selecting the right action based on partial information about the user.

%Ad platforms operate in environments where advertising decisions must be made under uncertainty about user preferences. Consider the case of an ad network deciding which ad to show to a particular user. The central goal of the platform is to maximize a reward measure, such as advertising revenue or user engagement. In practice, these objectives are closely tied to user clicks, since revenue and engagement are typically generated through click-based pricing and interaction mechanisms. At the same time, the platform does not fully observe who the user is or how they will respond to a given ad. As a result, the platform faces a decision-making problem under uncertainty, where the challenge lies in selecting the right action based on partial information about the user.

To analyze this decision problem, we model ad allocation as a choice among feasible actions in the presence of heterogeneous users and incomplete information. Users differ in latent preferences that determine the rewards generated by different ads, while the platform observes only an information set \(X\) that shapes its beliefs about these rewards. We now formalize the elements of this problem.

\begin{itemize}
\item \textbf{Environment.} The platform must choose an action from a finite set $\mathcal{A}$, where each $a \in \mathcal{A}$ represents a feasible option (e.g., an ad that can be shown). Users differ in unobserved characteristics captured by a latent type $\omega \in \Omega$, with $\Omega$ denoting the space of possible user types. The reward of taking action $a$ for a user of type $\omega$ is represented by a function:
\[
r: \mathcal{A} \times \Omega \to \mathbb{R}, \qquad r(a, \omega) \text{ (e.g., expected CTR)}.
\]
This reward function encodes the payoff of each action for each user type and serves as the platform’s objective.

\item \textbf{Information set and user types.} The platform does not observe the user's type $\omega$ directly. Instead, they observe an information set $X \in \mathcal{X}$ that provides partial knowledge about $\omega$. The information set $X$ may include behavioral variables, device attributes, geographical indicators, or other covariates. We assume the pair $(X, \omega)$ is drawn from a joint distribution $p$ over $\mathcal{X} \times \Omega$, which factorizes as \(p(x,\omega) = p(\omega)\,p(x\mid \omega)\), where $p(\omega) \in \Delta(\Omega)$ represents the population-level distribution of user types. Upon observing $X = x$, the platform forms a posterior belief over types:
\[
q_x(\omega) := p(\omega \mid x) \in \Delta(\Omega).
\]
For notational clarity, we also define the \emph{random posterior} $q_X := p(\cdot \mid X)$, a random variable that maps each realization $X = x$ to its corresponding posterior $q_x$. This posterior encodes the platform’s updated belief about the user's type after observing $x$.

\item \textbf{Policies.} A policy specifies how the platform selects actions based on the observed information set. Formally, a policy is a measurable function:
\[
\pi: \mathcal{X} \to \mathcal{A},
\]
which maps each realization $x \in \mathcal{X}$ to an action $\pi(x) \in \mathcal{A}$. We focus on deterministic policies because, under linear expected reward objectives, randomization offers no additional value (by an extreme-point argument; see \citealp{kamenica2019bayesian, smith2023optimal})\footnote{When $\mathcal{A}$ is finite and the objective is linear, any mixed policy can be written as a convex combination of deterministic ones, and the maximum is achieved at an extreme point. Ties are resolved by an arbitrary but fixed rule.}.

\item \textbf{Expected reward and optimal actions.}  
Given a belief $q \in \Delta(\Omega)$ over user types, such as the posterior $q_x$ induced by observing $x$, the expected reward of choosing action $a \in \mathcal{A}$ is  
\[
\mathbb{E}_{\omega \sim q_x}[r(a, \omega)].
\]  
Following, among all possible policies $\pi \in \Pi$ that map observed information to actions, the optimal policy $\pi^*(x)$ is the one that maximizes expected reward for each observation:
\begin{equation}\label{equ:optimalpolicy}
\pi^*(x) \in \arg\max_{a \in \mathcal{A}} \ \mathbb{E}_{\omega \sim q_x}[r(a, \omega)].
\end{equation}

\item \textbf{Decision value.} Let $X$ denote the information set observed prior to action selection, and let $q_X = p(\cdot \mid X)$ be the induced posterior over user types. For any deterministic policy $\pi: \mathcal{X} \to \mathcal{A}$, its ex-ante value is \(V(\pi)
= \mathbb{E}_{(X,\omega)\sim p}[r(\pi(X), \omega)]\). Therefore, the value of information from $X$ is the maximum achievable value over all such policies:
\begin{equation}\label{equ:valueFunction}
V_X = \sup_{\pi: \mathcal{X} \to \mathcal{A}} V(\pi)
= \sup_{\pi: \mathcal{X} \to \mathcal{A}} \mathbb{E}_{X}\big[\mathbb{E}_{\omega \sim q_X}[r(\pi(X), \omega)]\big].
\end{equation}
The optimal policy $\pi^*(x)$ defined in Equation \ref{equ:optimalpolicy} attains this supremum by maximizing expected reward for each observation.
\end{itemize}

In this study, we focus on click outcomes (CTR) as the primary measure of reward. Clicks are central to digital advertising because they are closely tied to both platform revenue and user engagement under prevalent cost-per-click and click-weighted pricing mechanisms. They provide an immediate and observable measure of user response that reflects the quality of the platform’s ad allocation decision. Given this setup, we operationalize rewards using click outcomes.\footnote{We focus on clicks as the primary outcome because targeting value arises from increasing the relevance and clickability of ads. However, the framework readily extends to alternative reward measures, such as revenue per click.}

Let $Y \in \{0,1\}$ denote the binary indicator of user engagement (e.g., whether the user clicks on the displayed ad), and let $a \in \mathcal{A}$ denote the action chosen by the platform. Consistent with the model above, the realized outcome $Y$ depends on both the selected action $a$ and the user’s latent type $\omega \in \Omega$, which captures unobserved preferences and motivations. While $\omega$ is not directly observable, the platform may observe proxy information $X$ that is informative about it. We categorize available user-level information into two main types:
\begin{itemize}
    \item \textbf{Behavioral Information} ($X^B \in \mathcal{X}^B$) captures a user’s past interactions in the digital environment, which includes prior app usage patterns, ad exposure sequences, click history, and other engagement-based indicators that reflect individual preferences and interests over time.
    \item \textbf{Geographical Information} ($X^G \in \mathcal{X}^G$) describes the user’s spatial context, encompassing attributes such as precise location (e.g., latitude/longitude) and broader administrative regions such as city/province that may correlate with demographic or regional characteristics.
\end{itemize}
We denote by $X^{GB}$ the information regime that combines both behavioral and geographical information. We also treat \emph{contextual information}, denoted by $X^{\varnothing}$, as standard metadata attached to each ad impression, including device characteristics (e.g., brand and operating system) and network conditions (e.g., connectivity type). Like behavioral and geographical information, contextual variables reflect aspects of the user’s latent type $\omega \in \Omega$, but they differ in two important respects. First, they are typically available to platforms for every impression, making them a natural baseline information set. Second, consumers generally view them as substantially less privacy-sensitive than detailed behavioral or location-based data \citep{jerath2024consumers}. Accordingly, we include contextual information ($X^{\varnothing}$) in all targeting regimes. The specific variables included in each regime $X \in \{X^{\varnothing}, X^{G}, X^{B}, X^{GB}\}$ are described in \S\ref{sssec:X}.

While we formalize the value of information through its impact on decision outcomes in Equation~\ref{equ:valueFunction}, a common alternative approach evaluates information solely through predictive performance. This approach focuses on estimating $\hat{y} = f(X,a)$ to approximate $\mathbb{P}(Y=1 \mid X,a)$ and assesses information quality using metrics such as AUC or reductions in conditional entropy $H(Y \mid X,a)$. However, predictive accuracy does not in general translate into better decisions or improved outcomes when targeting policies are deployed \citep{ascarza2018retention, rafieian2023ai}.

%\subsection{Challenges}\label{ssec:challenges}

\section{Empirical Framework}\label{sec:strategy}

The central objective of this framework is to assess the value of information by quantifying and comparing how behavioral data ($X^B$) and geographical data ($X^G$) contribute to improved decision-making. The value function in Equation~\ref{equ:valueFunction} provides the formal basis for this assessment by defining the maximum expected reward attainable under each information set. Implementing this framework, however, raises several challenges:

\begin{enumerate}
    \item \textbf{Challenge 1: Information Value, Complementarity, and Substitutability:} The first challenge is to formalize how to compare the value of distinct sources of information. When multiple data sources, such as behavioral (\(X^B\)) and geographical (\(X^G\)), are available, their value is inherently \emph{relative}: each may add incremental benefit beyond the other or overlap in what it conveys. The marginal value of one source depends on whether the other is already observed, raising the possibility of complementarity or substitutability. Without precise definitions, we cannot separate the standalone contribution of each source from the value created by using them jointly. We therefore develop a formal framework that isolates individual value, incremental value conditional on the other source, and their combined effect. We present this framework in \S\ref{ssec:sub_comp}.

    \item \textbf{Challenge 2: Reward Function Estimation for Decision Policies:} The second challenge is empirical: estimating the reward function \( r(a,\omega) \). Our objective is to assess whether geographical information provides incremental decision value beyond behavioral information. Doing so requires reward estimation that fully captures the dynamic structure of user histories; otherwise, estimated gains from geographical information may reflect behavioral misspecification rather than informational value. Because behavioral information is inherently temporal, reward estimation must account for this dynamic structure. We address this challenge in \S\ref{ssec:ML}. 

    \item \textbf{Challenge 3: Information Value Estimation through Policy Evaluation:} The third challenge concerns estimating the value of information \(V_X\). By definition, \(V_X\) is the maximum expected reward achievable under a given information set, which requires comparing alternative decision policies. Such comparisons are inherently counterfactual, since each user is observed under only one realized action and outcomes under other actions are not observed. Estimating \(V_X\) therefore requires a model-free approach to evaluating the performance of counterfactual policies using logged data, even when some actions are rarely chosen. We address this challenge in \S\ref{ssec:ips_estimation}.

\end{enumerate}

To address these challenges, we develop a unified framework that integrates economic theory, machine learning, and causal inference. The framework (i) formally defines information value to compare behavioral and geographical data and characterize complementarity or substitutability (\S\ref{ssec:sub_comp}), (ii) estimates reward functions that capture the dynamic structure of behavioral histories (\S\ref{ssec:ML}), and (iii) evaluates information value through model-free counterfactual policy evaluation using logged data (\S\ref{ssec:ips_estimation}). Together, these components quantify the decision value of behavioral and geographical information.

\subsection{Substitutability and Complementarity of Information Sets}\label{ssec:sub_comp}

To address Challenge 1, we use formal definitions from the economic theory literature to compare the value of different information sets in a decision problem. In our context, we assess how behavioral data \(X^B\) and geographical data \(X^G\) contribute to guiding optimal actions. Intuitively, two pieces of information complement each other if the joint incrementality in information value goes beyond a simple sum of the incrementality of each alone. %about the user’s latent type $\omega$. \mohammad{re-written}

We adopt the decision-theoretic framework of substitutability and complementarity introduced by \citet{borgers2013signals}. Recall that \(X^B\) and \(X^G\) denote behavioral and geographical data, respectively. We define the combined information set as \(X^{GB} := (X^B, X^G) \in \mathcal{X}^B \times \mathcal{X}^G\), which represents joint access to both sources. As a baseline, we consider \(X^{\varnothing}\), corresponding to the absence of user-level information. In this case, decisions rely only on \emph{contextual information} that informs the prior belief about user types, so the posterior reduces to the prior, \(q_{X^{\varnothing}} \equiv p(\omega)\). The associated value, \(V_{X^{\varnothing}}\), therefore represents decision-making under prior uncertainty without conditioning on individual-level signals. We define substitutability and complementarity by the marginal value of one information set conditional on access to the other, using the value function \(V_X\) (Equation~\ref{equ:valueFunction}).

\begin{definition}[Substitutes]\label{def:Sub}
Behavioral information \(X^B\) is a \emph{substitute} for geographical information \(X^G\)
\emph{if, for every decision problem \footnote{Following \citet{borgers2013signals}, “every” refers to the inequalities holding for \emph{all} decision problems \((A,r)\). Our empirical analysis evaluates these same inequalities for a fixed problem \((A,r)\) (CTR objective and action set), yielding complement/substitute conclusions \emph{for this problem}.} \((\mathcal A,r)\)
,}
\[
  V_{X^G} - V_{X^\varnothing} \;\ge\; V_{X^{GB}} - V_{X^B}.
\]
\end{definition}

This inequality compares the marginal value of geographical data when used alone versus when combined with behavioral data. The left-hand side, \( V_{X^G} - V_{X^\varnothing} \), reflects the value of geographical data on its own, while the right-hand side, \( V_{X^{GB}} - V_{X^B} \), reflects its incremental contribution when behavioral data are already available. If the inequality holds, behavioral data reduces the marginal usefulness of geographical data, making the two substitutes.

\begin{definition}[Complements]\label{def:Comp}
Behavioral information \(X^B\) is a \emph{complement} to geographical information \(X^G\)  
\emph{if, for every decision problem \((\mathcal{A},r)\),}
\[
  V_{X^G} - V_{X^\varnothing} \;\le\; V_{X^{GB}} - V_{X^B}.
\]
\end{definition}
Here, the marginal value of geographical data is greater when combined with behavioral data than when used alone. The left-hand side, \(V_{X^{GB}} - V_{X^B}\), measures the incremental contribution of geographical data given that behavioral data are already available, while the right-hand side, \(V_{X^G} - V_{X^\varnothing}\), captures their standalone value. If the inequality holds, behavioral and geographical data enhance each other’s usefulness and are thus considered complements.

\subsection{Machine Learning Framework for Reward Function Estimation}\label{ssec:ML}

To address Challenge 2, we develop a machine learning framework for estimating expected reward when the reward function \( r(a, \omega) \) is unobserved. The ad platforms cannot observe latent types \( \omega \), nor directly measure the reward associated with each action. Instead, we re-express reward in terms of observable data: user characteristics \(X\), chosen actions \(A\), and realized engagement outcomes \(Y \in \{0,1\}\).  

We specify reward through the binary click outcome \(Y\) and the reward value of engagement. Suppose that showing ad \( a \in \mathcal{A} \) to a user of type \( \omega \in \Omega \) yields reward only when the user clicks, and that the ad platform receives a known per-click reward \( v(a) \in \mathbb{R}_{\geq 0} \).\footnote{Since the targeting value of information arises from improved match quality and resulting changes in users’ CTR, we focus on this reward measure and set $v(a)=1$ for simplicity. All insights extend to settings in which $v(a)$ is heterogeneous, provided that CTR and per-click valuation are independent and separable.} In this case, the expected reward from taking action \( a \) for a user of type \( \omega \) is given by  
\[
r(a, \omega) := \mathbb{E}[Y \mid A = a, \omega] \cdot v(a) = \Pr(Y = 1 \mid A = a, \omega) \cdot v(a).
\]
This formulation captures both the probabilistic nature of engagement and the economic payoff per click, and it serves as the foundational definition of reward in our analysis. Although user types \( \omega \) are unobserved, conditioning on observed characteristics \( X = x \) induces a posterior belief \( q_x(\omega) := p(\omega \mid x) \) about the user’s latent type. Accordingly, we focus on estimating the function:  
\[
f_X(a, x) := \Pr(Y = 1 \mid X = x, A = a),
\]
which maps each action–characteristics pair to the probability of engagement. This function is the central object in our machine learning framework. Once \( f_X \) is estimated, the expected reward can be easily computed. This approach can be operationalized using flexible prediction models trained on observational data, provided that the model class is sufficiently expressive and the estimation procedure is properly regularized and calibrated. 

%Once \( f_X \) is estimated, the expected reward can be computed as:
%\[
%\widehat{\mathbb{E}}\!\left[r(a,\omega)\mid X=x\right]
%= v(a)\cdot \hat f_X(a,x).
%\]
%and the ad platform can select actions by maximizing this estimated utility. This approach can be operationalized using flexible prediction models trained on observational data, provided that the model class is sufficiently expressive and the estimation procedure is properly regularized and calibrated. 

To operationalize this framework, we estimate the conditional click function under several information regimes that differ in the richness of observable user data. Specifically, we consider four regimes \( \{X^\varnothing, X^G, X^B, X^{GB}\},\) which correspond to progressively richer access to contextual, geographical, and behavioral information. Each regime captures a distinct informational constraint faced by the advertising platform. The specific features included in each information set are described in Appendix~\S\ref{sssec:X}. 

In the following, we describe learning algorithm selection \( f \) from a flexible class \( \mathcal{F} \) to approximate \( \Pr(Y = 1 \mid X, A) \) in \S\ref{sssec:MLselection}, and we specify the behavioral modeling architecture in \S\ref{sssec:ML_Arch}. The estimated function \( \hat{f}_X \) delivers action-specific rewards, induces policies \( \pi(x) \), and forms the basis for policy evaluation across informational regimes.

\subsubsection{Learning Algorithm Selection Across Information Regimes}\label{sssec:MLselection}
In regimes without behavioral histories (\(X^\varnothing\) and \(X^G\)), the information set is static across impressions. Because no new user-specific information accumulates over time, the estimation problem is inherently cross-sectional. We therefore employ flexible tree-based learners (XGBoost) to estimate the click probability as a function of contemporaneous covariates and actions. These models are well-suited to capturing nonlinearities and high-order interactions in static feature spaces without imposing strong parametric structure \citep{Rafieian2021}.

By contrast, regimes that include behavioral information (\(X^B\) and \(X^{GB}\)) generate sequential user histories whose informational content expands over time. In these settings, the relevant state variable is no longer a fixed vector but an evolving sequence of past exposures and engagement outcomes. To accommodate this structure, we develop a sequence model based on the LSTM network with an attention mechanism, which aggregates information over time and forms latent summaries of user behavior \citep{quadrana2018sequence}. These models map behavioral histories and current actions into predicted engagement probabilities, allowing the estimator to adapt dynamically as additional impressions are observed. %Details on model selection are provided in Appendix~\S\ref{sssec:MLselection}, and the architecture of the sequential LSTM model is described in Appendix~\S\ref{sssec:ML_Arch}.

This distinction between static and sequential regimes is central to our analysis. As behavioral information accumulates over impressions, the information set \(X^B_{jt}\) becomes increasingly informative about user behavior. This richer information allows the advertiser to condition decisions on a more precise representation of past engagement, improving estimation of the engagement function \(f(X^B_{jt}, a)\). As a result, targeting decisions become more accurate, and the expected reward under the induced policy increases as behavioral histories grow. We examine this empirically in \S\ref{ssec:HeterogeneityValue}.

\subsubsection{Sequence-Based Behavioral Modeling with LSTM and Attention}\label{sssec:ML_Arch}
We now describe the sequence model used for regimes that include behavioral histories in this part. We design an LSTM architecture equipped with a causal multi-head attention mechanism. The LSTM captures short- and long-range temporal dependencies in user interactions, while attention highlights the most informative parts of a user’s exposure–click history. Together, these components allow the model to represent both persistent behavioral trends and short-term recency effects, which are central to engagement modeling \citep{quadrana2018sequence}.

We train the model on user interaction histories using a sliding window of 150 impressions. At each time step, the input combines categorical and numerical features. We map high-cardinality categorical variables (e.g., device model, app ID, network identifiers) into dense vectors with embedding layers \citep{guo2016entity} and concatenate them with continuous covariates. To encode order, we add absolute positional embeddings \citep{vaswani2017attention}. We also process the log-transformed inter-arrival time through a time-gap projection, which helps the model detect irregular timing patterns that often indicate behavioral shifts.

\begin{figure}[htp!]
    \centering
    \includegraphics[width=0.75\linewidth]{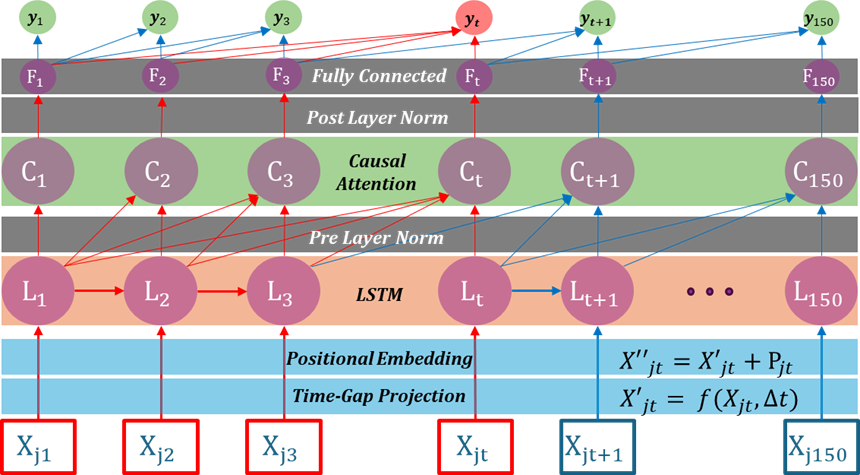}
    \caption{LSTM with causal attention architecture}
    \label{fig:lstm_architecture}
\end{figure}
We feed the enriched sequence into a four-layer LSTM with hidden size 512. On top of the LSTM outputs, we leverage causal multi-head self-attention with four heads. The attention module uses a future-masked matrix to block access to unseen impressions, which preserves temporal causality and prevents information leakage \citep{vaswani2017attention}. We apply layer normalization before and after the attention block to stabilize training. Next, we add a gated projection head that runs parallel sigmoid and tanh transformations, combines them elementwise, and applies dropout for regularization. A fully connected output layer finally maps the hidden states to click probabilities at each step.

Figure~\ref{fig:lstm_architecture} illustrates the designed sequence model architecture and its key components. The design directly leverages sequential order, irregular timing, and varying relevance of past impressions. As users generate longer histories, the LSTM–attention framework provides richer information about preferences and latent types. We provide additional implementation details in Web Appendix \S\ref{app:LSTM}.

\subsection{Estimating the Value of Information via Inverse Propensity Scoring}\label{ssec:ips_estimation}
Challenge 3 highlights the core difficulty in estimating the value of information $V_X$(Equation~\ref{equ:valueFunction}): 
\[
V_X = \sup_{\pi: \mathcal{X} \to \mathcal{A}} \mathbb{E}_X\left[\mathbb{E}_{\omega \sim q_X}[r(\pi(X), \omega)]\right],
\]
The data only reveal outcomes under the action chosen by the platform’s logging policy, while $V_X$ requires utilities aggregated over all possible actions. Each impression, therefore, provides information on one action, but evaluating a policy involves counterfactual outcomes for actions that were not taken.

We address this problem with Inverse Propensity Scoring (IPS), a standard method in causal inference for adjusting observational data \citep{hirano2003efficient, rafieian2023ai}. IPS identifies the value of a fixed target policy by reweighting observed outcomes according to how likely the logging policy was to select the same action as the target policy. By doing so, we approximate the reward that would have been realized under alternative policies, even though those policies were never deployed in practice.

Formally, for each information set $X$, we define a target policy $\pi^{f_X}$ that deterministically selects the action maximizing the estimated expected reward based on our predictive model:
\[
\pi^{f_X}(a \mid x) \;=\; \mathbb{I}\!\left\{a \;=\; \arg\max_{a' \in \mathcal{A}} \hat{f}_X(a, x)\right\}, 
\]
where $\hat{f}_X(a, x)$ is the estimated click probability given information set $x$ and candidate action $a$. Thus, $\pi^{f_X}$ prescribes, for each instance, the action expected to maximize reward under the model. The value of this policy, consistent with our decision-theoretic objective, is

\[
V(\pi^{f_X}) \;=\; \mathbb{E}_X\!\left[\mathbb{E}_{\omega\sim q_X}\big[r(\pi^{f_X}(X),\omega)\big]\right].
\]
However, only outcomes corresponding to the logging policy’s actions are observed in the data. To estimate the value of the induced policy $V(\pi^{f_X})$, we use the IPS estimator, which reweights the observed reward outcomes for each user-impression pair by the inverse of the probability that the logging policy would have selected the action recommended by the target policy:
\[
\hat V(\pi^{f_X})
= \frac{1}{N}\sum_{i=1}^N 
\frac{\mathbb{I}\!\big[A_i=\pi^{f_X}(X_i)\big]}{\pi^{\mathcal D}(A_i\mid X_i)} \, v(A_i)\, Y_i,
\]
where $A_i$ is the action taken in the historical data for impression $i$, $\pi^{\mathcal{D}}(A_i \mid X_i)$ is the probability that the logging policy selected $A_i$ given information $X_i$, and $Y_i$ is the observed binary engagement outcome. Each term, therefore, reweights the realized reward to reflect how the target policy would allocate actions. For the IPS estimator to recover the true value $V_X$, three key conditions must hold regarding the data-generating process and the structure of the logging policy. We state these assumptions formally below.

\begin{assumption}[Overlap]\label{ass:overlap}
For all $x \in \mathcal{X}$ and $a \in \mathcal{A}$, if $\pi^{f_X}(a \mid x) > 0$, then $\pi^{\mathcal{D}}(a \mid x) > 0$. In other words, the logging policy must assign positive probability to every action that the target policy might select.
\end{assumption}

\begin{assumption}[Unconfoundedness]\label{ass:unconfoundedness}
Potential outcomes are independent of the action actually taken, conditional on observed covariates $X$; that is, $Y(a) \perp A \mid X$ for all $a \in \mathcal{A}$. This ensures that all confounding factors are captured in $X$.
\end{assumption}

\begin{assumption}[Policy optimality under $X$]\label{ass:optimal}
If $\hat f_X$ consistently estimates $\Pr(Y=1\!\mid\!A=a,X=x)$ and the maximizer of $v(a)\Pr(Y=1\!\mid\!A=a,X=x)$ is unique almost everywhere, then $\pi^{f_X}(x)=\arg\max_a v(a)\hat f_X(a,x)$ is optimal under $X$ almost everywhere.
\end{assumption}

\begin{proposition}[Standard IPS identification and recovery of $V_X$]\label{prop:ipsVX}
Under Assumptions~\ref{ass:overlap}--\ref{ass:optimal}, and when $\pi^{f_X}$ is evaluated on an independent sample,
\[
\mathbb{E}\!\big[\hat V(\pi^{f_X})\big] \;=\; V_X,
\qquad 
\hat V(\pi^{f_X}) \xrightarrow{p} V_X.
\]
That is, the IPS estimator is unbiased and consistent for $V_X$ under the stated conditions. \\  
\begin{proof}
A formal proof is provided in Web Appendix \S\ref{app:ips_proof}.
\end{proof}
\end{proposition}

The IPS approach addresses Challenge 3 by enabling estimation of the ex-ante value of any information set \( X \) through the policy it induces, even when counterfactual outcomes are unobserved. In the following, we first discuss the conditions and diagnostics required to validate propensity score estimation in \S\ref{sssec:auction-ips}, and then describe the empirical procedure used to estimate propensity scores in \S\ref{sssec:empiricalpropensity}.

\subsubsection{Estimating and Validating Propensity Scores }\label{sssec:auction-ips}

To implement the IPS estimator, we require estimates of the logging policy probabilities \( \pi^{\mathcal{D}}(a \mid x) \), which govern how actions are assigned in the historical data. First, we describe the ad allocation mechanism that generates these probabilities, and then examine empirical evidence supporting the IPS identification assumptions.

\paragraph{Quasi-Proportional Allocation Mechanism.}
In our setting, impressions are allocated via a \emph{quasi proportional auction} that induces randomized exposure across eligible ads as a function of observed bids, quality scores, and eligibility constraints \citep{mirrokni2010quasi}. For each impression \( i \), let \( \mathcal{A}_i \subseteq \mathcal{A} \) denote the set of ads eligible to participate in the auction. Each ad \( a \in \mathcal{A}_i \) submits a bid \( b_a \) and has a platform-assigned quality score \( q_a \), both of which are fixed and observed during our sample period. The probability that ad \( a \) wins the auction and is shown in impression \( i \), conditional on covariates \( x_i \), is given by:
\[
\pi^{\mathcal{D}}(a \mid x_i) = \frac{b_a q_a}{\sum_{j \in \mathcal{A}_i} b_j q_j}.
\]
This allocation rule induces a probabilistic assignment over eligible ads, with probabilities that are fully determined by observed features. In contrast to deterministic formats (e.g., second-price auctions) where only the top-ranked ad is observed, the quasi-proportional mechanism introduces randomization across all eligible ads. This variation is central to our empirical framework, as it enables estimation of counterfactual outcomes and supports off-policy evaluation.

\begin{remark}[Overlap]
Any ad participating in the auction for impression \( i \) (i.e., \( \forall a \in \mathcal{A}_i \)) has a nonzero propensity of being shown in impression \( i \).
\end{remark}
This follows directly from the quasi-proportional rule: every ad with a positive \( b_a q_a \) has a strictly positive probability of being selected. Hence, the overlap assumption \ref{ass:overlap} is satisfied by construction for all participating ads in each auction.

\begin{remark}[Unconfoundedness]
For any impression \( i \), ad allocation is independent of the set of potential outcomes for participating ads \( (a \in \mathcal{A}_i) \), after controlling for the observed covariates. Thus, 
\[
\{Y_i(a)\}_{a \in \mathcal{A}_i} \perp A_i \mid x_i.
\]
\end{remark}
Therefore, the unconfoundedness assumption \ref{ass:unconfoundedness} is satisfied by the transparent structure of the auction: all inputs that determine allocation, namely, bids \( b_a \), quality scores \( q_a \), and eligibility constraints, are fully observed and fixed over the sample period. For each impression \( i \), we observe the complete covariate vector \( x_i \). Advertiser bids are not dynamically adjusted, and the platform does not personalize quality scores across users. As a result, the assignment rule is fully determined conditional on \( x_i \).

This structure implies that for any impression \( i \), we can compute not only the probability \( \pi^{\mathcal{D}}(A_i \mid x_i) \) of the ad that was actually shown, but also the probabilities of all counterfactual ads that were eligible in the same auction. That is, even if a particular ad \( a \) was not displayed in impression \( i \), as long as it was eligible (i.e., \( a \in \mathcal{A}_i \)), we can recover the counterfactual allocation probability \( \pi^{\mathcal{D}}(a \mid x_i) \). 

\paragraph{Eligibility Filtering.}
Although the quasi-proportional rule defines probabilities \( \pi^{\mathcal{D}}(a \mid x_i) \) for all ads in the auction, not all ads in the global set \( \mathcal{A} \) are eligible in every impression. To ensure valid counterfactual estimation, we restrict attention to ads with nonzero probability of participation in each impression. Two factors determine eligibility: (i) \textit{Contextual Targeting}: ads may be restricted to specific provinces, times, or app categories, and are excluded when their targeting criteria are not met; (ii) \textit{Campaign availability}: some ads may be inactive due to budget exhaustion or campaign timing. In practice, this is rare among top ads, as we select the top 10 ads for our analysis.

We construct an eligibility matrix \( E \in \{0,1\}^{N \times A} \), where \( e_{i,a} = 1 \) indicates that ad \( a \) was eligible to compete in impression \( i \), based on observed targeting and availability constraints. In practice, this requires that the impression’s metadata match the ad’s targeting filters on province, hour-of-day, and app. To avoid misclassification, we drop any impressions with missing targeting variables and restrict attention to a filtered sample where eligibility can be verified.

While this filtering step identifies the support of the logging policy, it does not tell us how likely each eligible ad is to be selected. For unbiased off-policy evaluation, we must account for the non-random assignment probabilities across eligible ads. Next, we describe how we estimate these propensities and assess the validity of the unconfoundedness assumption via covariate balance.

\subsubsection{Propensity Score Estimation and Covariate Balance}\label{sssec:empiricalpropensity}
To correct for unequal selection probabilities inherent in the auction, we estimate the \emph{propensity scores} 
\(\pi^{\mathcal{D}}(a \mid x_i)\), defined as the probability that the logging policy assigns impression \(i\) to ad \(a\) given the observed features \(x_i\). These propensities quantify the exposure pattern generated by the platform’s allocation mechanism and form the basis for IPS in our policy evaluation.

Although the quasi-proportional rule maps bids and quality scores into theoretical probabilities, we adopt a data-driven estimation strategy to capture the realized assignment process in practice. This approach flexibly accommodates deviations from the theoretical rule, nonlinear effects, and high-order interactions among features. We use XGBoost to estimate \(\pi^{\mathcal{D}}(a \mid x_i)\), motivated by its empirical performance in high-dimensional classification problems \citep{rafieian2023ai}. To avoid overfitting and ensure out-of-sample validity, we implement a 5-fold cross-fitting procedure, so that each estimated propensity score is computed on a model trained without the corresponding observation. 

While observed inputs fully determine the allocation rule, we validate our identifying assumption by testing whether inverse-propensity weighting balances the distribution of observed covariates across treatment. The variables examined include province, app context, time of day, device brand, network type, and mobile service provider, features that advertisers can directly target. For each covariate and treatment group, we compute the standardized mean difference (SMD) before and after applying the estimated propensity weights, considering absolute values below 0.2 as indicative of acceptable balance \citep{mccaffrey2013tutorial}. The diagnostics show substantial improvements in covariate alignment across ads after weighting, providing empirical support for our research design. Details of propensity score estimation and balance statistics are reported in Web Appendix \S\ref{app:propensity_estimation}.

\section{Empirical Results}\label{sec:Results}
We now present the empirical results derived from our proposed framework.  
\S\ref{ssec:MLPerformance} reports the predictive performance of machine learning models trained on different information sets. We then turn to the core question of how behavioral and geographical information contribute to decision quality, examining it at two levels of analysis. First, at the \emph{aggregate level} (\S\ref{ssec:InformationValue}), we quantify the overall value of each information set and test whether the two act as substitutes or complements in improving targeting performance. Second, we extend the analysis to the \emph{user level} (\S\ref{ssec:HeterogeneityValue}), examining how the value and interaction of these information types change with the amount of behavioral history observed by the platform.

\subsection{Predictive Performance Across Information Sets}\label{ssec:MLPerformance}
Predicting user engagement is inherently challenging due to extreme class imbalance: fewer than 2\% of impressions generate a click, making accuracy an uninformative metric. We therefore evaluate predictive performance using log loss, which assesses the calibration of predicted click probabilities and aligns with the models’ training objective; relative information gain (RIG), which measures the proportional improvement in log loss relative to a baseline model that predicts the average click-through rate; and the area under the ROC curve (AUC), which captures the model’s ability to rank clicked above non-clicked impressions in a threshold-independent and imbalance-robust manner.

We evaluate four models trained under different informational regimes: (i) the contextual-only baseline (\(X^{\varnothing}\)), (ii) the geographical regime (\(X^G\)), (iii) the behavioral regime (\(X^B\)), and (iv) the full information set (\(X^{GB}\)). The performance of each model on the held-out test set is summarized in Table~\ref{tab:auc_rig}.

\begin{table}[htp!]
\centering
\begin{threeparttable}
\caption{Predictive performance of models using contextual data ($X^{\varnothing}$), geographical data ($X^G$), behavioral data ($X^B$), and both ($X^{GB}$).}
\label{tab:auc_rig}
\begin{tabular}{lccc}
\toprule
\textbf{Model} & \textbf{Log Loss} & \textbf{AUC} & \textbf{Relative Information Gain (\%)} \\
\midrule
($X^{\varnothing}$) & 0.073 & 0.720 & 18.900 \\
($X^G$)             & 0.072 & 0.722 & 19.220 \\
($X^B$)             & 0.014 & 0.809 & 84.180 \\
($X^{GB}$)          & 0.013 & 0.812 & 84.410 \\
\bottomrule
\end{tabular}
\begin{tablenotes}[para,flushleft]
\footnotesize
Notes: The table reports Log Loss, AUC, and RIG on the test set. RIG is computed relative to the baseline click-through rate $p=0.017592$. Sample size is $N=3{,}162{,}376$ impressions.
\end{tablenotes}
\end{threeparttable}
\end{table}

The results reveal substantial variation in predictive performance across information structures. The contextual-only baseline performs weakest, with an RIG of 18.90\%. Adding geographical features yields only a modest improvement, raising RIG to 19.22\% and AUC to 0.722. In contrast, incorporating behavioral information leads to a large performance gain: the behavioral model achieves an RIG of 84.18\% and an AUC of 0.809, indicating both substantially richer information and a superior ability to rank click outcomes. Adding geographical information to the behavioral model produces only marginal additional improvements, suggesting that behavioral features account for most of the predictive variation.

Taken together, these results demonstrate the dominant predictive value of behavioral information in estimating click likelihood, which aligns with prior findings in the advertising  \citep{Rafieian2021}. While comparable AUC levels can be achieved using XGBoost-based behavioral models, the LSTM architecture delivers substantially higher RIG, indicating improved probabilistic calibration through temporal modeling. These performance levels provide empirical support for Assumption~\ref{ass:optimal}. Additional comparisons between LSTM and XGBoost models are reported in Appendix~\S\ref{app:XGboostLSTM_Prediction}.

That said, it is important to emphasize again that predictive performance does not always translate into decision value. A model may be well-calibrated or rank instances correctly, yet offer limited benefit when used to guide actions under realistic constraints. We therefore turn next to assessing the actual targeting value generated by each model in terms of the value function.

\subsection{Behavioral vs.\ Geographical Information: Aggregate Level}\label{ssec:InformationValue}
Having established model performance in \S\ref{ssec:MLPerformance}, we now assess how behavioral and geographical information improve decision quality. As described in \S\ref{ssec:ips_estimation}, we evaluate each targeting policy \( \pi^{(f_X)} \) using IPS. We proceed in two steps. First, in \S\ref{sssec:value-overall}, we quantify the \emph{aggregate value} of each information set by comparing the expected value of policies based on different data sources.  
Then, in \S\ref{sssec:ComSubG}, we test whether behavioral and geographical information act as \emph{substitutes or complements} in shaping decision quality.

\subsubsection{Aggregate Value of Behavioral and Geographical Information}\label{sssec:value-overall}

We consider a set of deterministic greedy policies indexed by the information set \( X \in \{X^{\varnothing}, X^G, X^B, X^{GB}\} \) available at the time of decision. For each impression \( i \), the policy selects, among eligible actions \( \mathcal{A}_i \) (see \S\ref{sssec:auction-ips}), the ad with the highest predicted click probability, restricting attention to actions with positive estimated logging propensity \( \hat{\pi}^{\mathcal D}(a \mid X_i) > 0 \). This ensures that all selected actions lie within the support of the logging policy and satisfy the overlap condition in Assumption~\ref{ass:overlap}.
%\footnote{Since the targeting value of information arises from improved match quality and resulting changes in users’ CTR, we focus on this reward measure and set $v(a)=1$ for simplicity. All insights extend to settings in which $v(a)$ is heterogeneous, provided that CTR and per-click valuation are independent and separable.}

Each policy represents a counterfactual targeting scenario in which the advertiser optimizes using only the corresponding information set \( X \), and its value \( \hat V_X \) is estimated via IPS. To assess estimator uncertainty, we report 95\% confidence intervals based on cluster-robust standard errors (clustered at the user level) and the effective sample size (ESS), which captures variance inflation due to skewed importance weights \citep{kallus2019stable}.

The results are reported in Table~\ref{tab:policy_value_ips}. Across all policies, we find statistically significant improvements over the historical logging baseline, with tight confidence intervals and high ESS. In particular, \citet{mccaffrey2013tutorial} advocates trimming weights until \(\mathrm{ESS} \geq 0.10N\), a threshold all policies in our setting easily exceed. This suggests that the estimated policy values are stable and inference is well-powered.

\begin{table}[htbp]
\centering
\caption{Estimated Policy Value Using IPS}
\label{tab:policy_value_ips}
\begin{tabular}{lcccccc}
\toprule
Policy & IPS Estimate & 95\% CI & t-stat & SE & Lift (\%) & ESS \\
\midrule
$\hat{V}_{X^{GB}}$        & 0.0248$^{***}$ & [0.0243, 0.0254] & 27.611 & 0.000 & 41.450 & 760316 \\
$\hat{V}_{X^B}$           & 0.0229$^{***}$ & [0.0223, 0.0234] & 19.675 & 0.000 & 30.280 & 555871 \\
$\hat{V}_{X^G}$           & 0.0226$^{***}$ & [0.0220, 0.0232] & 16.572 & 0.000 & 28.710 & 592427 \\
$\hat{V}_{X^\varnothing}$ & 0.0208$^{***}$ & [0.0202, 0.0214] & 18.782 & 0.000 & 20.440 & 542747 \\
\bottomrule
\end{tabular}
\\[1.ex]
\begin{minipage}{0.9\textwidth}
\footnotesize
\textit{Notes:} All estimates are computed using IPS. The baseline CTR under the logging policy is 0.017592. Lift is computed relative to this baseline. SE denotes the standard error of the estimate. Effective sample size (ESS) measures the number of equally weighted observations that would yield equivalent precision.  
Number of observations: 3,162,376. Cluster-robust standard errors are computed using 141,595 user clusters.  
$^{*}p<0.05$, $^{**}p<0.01$, $^{***}p<0.001$
\end{minipage}
\end{table}

Starting from the contextual information policy \( X^\varnothing \), we observe a 20.4\% improvement over the baseline CTR. Although the platform lacks access to user-level data, this policy leverages contextual variation and selects the ad with the highest average CTR. This result highlights the value of exploiting aggregate performance differences even in the absence of personalized information.

Access to richer information sets produces further gains in policy value. Geographical information (\( X^G \)) and behavioral information (\( X^B \)) both generate substantial improvements relative to the contextual policy, increasing lift to 28.7\% and 30.3\%, respectively. Combining both information sources (\( X^{GB} \)) yields the highest policy value, a 41.5\% improvement over the logging baseline. Notably, unlike the predictive results in \S\ref{ssec:MLPerformance}, where behavioral information explained most of the performance gain, geographical information contributes comparably to decision value. This pattern underscores that improvements in predictive accuracy do not necessarily translate into proportional improvements in decision quality. Additional comparisons across learning algorithms are reported in Web Appendix~\S\ref{app:XGboostLSTM_Value}.

\subsubsection{Complement or Substitute? Aggregate Level}\label{sssec:ComSubG}
We now investigate whether behavioral and geographical data act as \emph{substitutes} or \emph{complements} in informing ad targeting decisions, starting with the aggregate level. Formally, let:
\[
\Delta \;=\; \big(\hat{V}_{X^{GB}} - \hat{V}_{X^B}\big) - \big(\hat{V}_{X^G} - \hat{V}_{X^\varnothing}\big),
\]
where $\hat{V}_{X}$ denotes the IPS-based value estimate for policy $X$. The first term in parentheses measures the incremental value of adding geographical information when behavioral data are already available, while the second term measures the incremental value of adding geographical information when no user-level data are observed. 

If $\Delta > 0$, behavioral and geographical information are \emph{complements} (Definition~\ref{def:Comp}), meaning the value of combining them exceeds the sum of their stand-alone gains. If $\Delta < 0$, they are \emph{substitutes} (Definition~\ref{def:Sub}), meaning the combined gain is less than additive and the two information sets overlap in the information they provide. We estimate $\Delta$ using per-impression differences and conduct a one-sample $t$-test on the mean, clustering standard errors at the user level to account for correlation within users. We report the two-sided test, as one-sided results cannot be significant when the two-sided test fails to reject the null.

\begin{table}[htbp]
\centering
\caption{Aggregate Complementarity Test for Behavioral and Geographical Information}
\label{tab:aggregate_comp_sub}
\begin{tabular}{lccccc}
\toprule
 & Estimate ($\hat{\Delta}$) & Std. Error (clustered) & $t$-stat & 95\% Confidence Interval & $p$-value \\
\midrule
Value & 0.000509 & 0.000334 & 1.526 & [-0.000144, 0.001164] & 0.127 \\
\bottomrule
\end{tabular}
\\[1ex]
\begin{minipage}{0.92\linewidth}
\footnotesize
\textit{Notes:} The table reports the aggregate test of complementarity between behavioral and geographical information.  
Standard errors are clustered at the user level. Number of observations = 3,162,376; number of user clusters = 141,595.  
\end{minipage}
\end{table}

As shown in Table~\ref{tab:aggregate_comp_sub}, the aggregate interaction estimate is close to zero and statistically insignificant. This implies that, on average, combining behavioral and geographical data produces nearly additive gains, with no systematic evidence of complementarity or substitutability at the aggregate level. While each information set captures distinct user heterogeneity, their joint effect does not amplify or diminish targeting value in the aggregate. Given this null aggregate finding, we next examine whether the nature of the interaction varies with user impression depth to assess whether heterogeneity in complementarity or substitutability emerges across users.

\subsection{Behavioral vs.\ Geographical Information: Heterogeneity by User Exposure}\label{ssec:HeterogeneityValue}
The null result at the aggregate level motivates a closer examination of how the value and interaction of behavioral and geographical information vary across users.  
We focus on heterogeneity by \emph{impression depth}, the number of impressions observed from each user, which proxies how much the platform has learned about each user. We proceed in two steps. First, in \S\ref{sssec:ValueHeterogeneity}, we assess how the decision value of each information set changes with user exposure. Then, in \S\ref{sssec:ComSubHeterogeneity}, we test whether the relationship between behavioral and geographical information shifts from substitutive to complementary as more behavioral data accumulate.

\subsubsection{Heterogeneous Value of Behavioral and Geographical Information by User Exposure}\label{sssec:ValueHeterogeneity}
Building on the aggregate results presented in \S \ref{sssec:value-overall}, we now explore how the value of information varies with the user's impression history. Specifically, we investigate whether the gain from targeting differs depending on how many impressions a user has previously seen. This analysis is motivated by the discussion in ~\S\ref{ssec:ML}, which posits that the estimation of reward function \( f(X^B_t, a) \) becomes more accurate as more behavioral data is observed. If true, the marginal value of behavioral or combined data may evolve with impression depth.

\begin{figure}[htbp]
\centering
\begin{subfigure}[t]{0.5\textwidth}
    \centering
    \includegraphics[width=\linewidth]{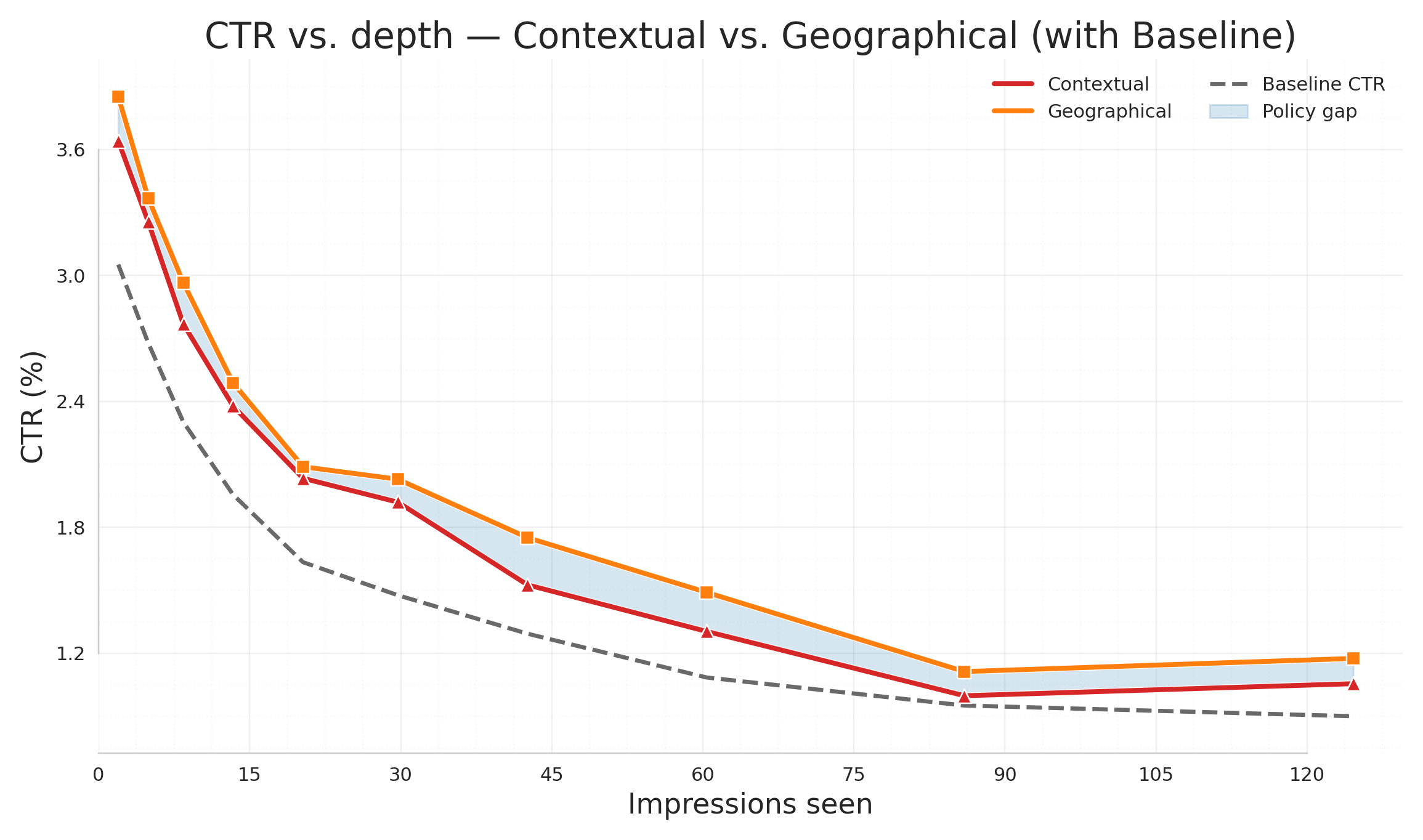}
    \caption{Contextual vs. Geographical Targeting.}
    \label{fig:lift_depth_C_G}
\end{subfigure}%
\hfill
\begin{subfigure}[t]{0.5\textwidth}
    \centering
    \includegraphics[width=\linewidth]{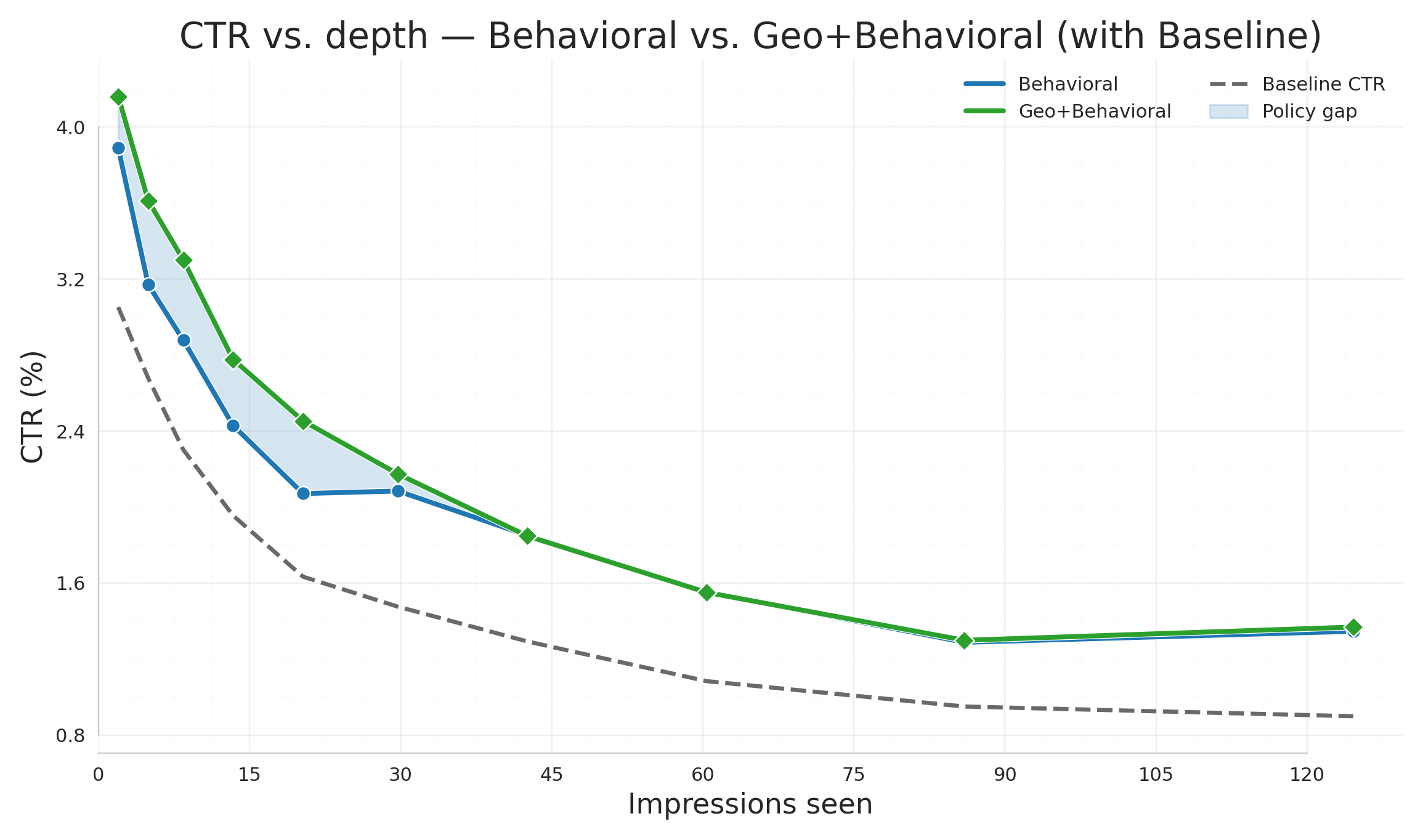}
    \caption{Behavioral vs. Geo + Behavioral Targeting.}
    \label{fig:lift_depth_B_GB}
\end{subfigure}
\vspace{0.5em}
\caption{Heterogeneity in Policy Value by Impression Depth. Each plot shows the \textit{absolute CTR levels} (in \%) for two targeting policies and the empirical baseline. Shaded areas indicate the policy gap, while the dashed line marks the baseline CTR.}

\label{fig:heterogeneity_plots}
\end{figure}

To operationalize this, we first sort all impressions for each user \( j \) by their timestamp \( t \) and assign a depth index accordingly. We then group observations into bins such that each bin contains the same number of impressions (i.e., quantile-based binning). Within each bin, we compute the absolute CTR levels for each targeting policy using the IPS-estimated click probabilities, alongside the empirical baseline CTR. These per-bin averages are then plotted against impressions have seen by user to visualize how click-through performance evolves as users receive additional exposures. Figure~\ref{fig:lift_depth_C_G} and Figure~\ref{fig:lift_depth_B_GB} present these results for the main targeting comparisons.

In the left panel of Figure~\ref{fig:lift_depth_C_G}, the \(X^G\) (Geographical) policy consistently delivers higher CTR than the \(X^\varnothing\) (Contextual) policy across all impression depths. This persistent gap indicates that location-based segmentation provides a stable improvement over contextual cues alone. Although both policies rely on static features, geographical information captures cross-regional heterogeneity that enhances targeting precision even in the absence of behavioral data.

In the right panel of Figure~\ref{fig:lift_depth_B_GB}, both the \(X^B\) (Behavioral) and \(X^{GB}\) (Combined) policies achieve substantially higher CTR than the baseline, highlighting the value of behavioral information for personalization. The \(X^{GB}\) policy consistently outperforms \(X^B\), though the gap narrows as impression depth increases. This pattern implies that while geographical data initially enhances personalization in early exposures, its marginal contribution diminishes once rich behavioral histories accumulate.

\subsubsection{Complement or Substitute? Heterogeneity by User Exposure}\label{sssec:ComSubHeterogeneity}

\begin{figure}[htp!]
    \centering
    \includegraphics[width=0.9\linewidth]{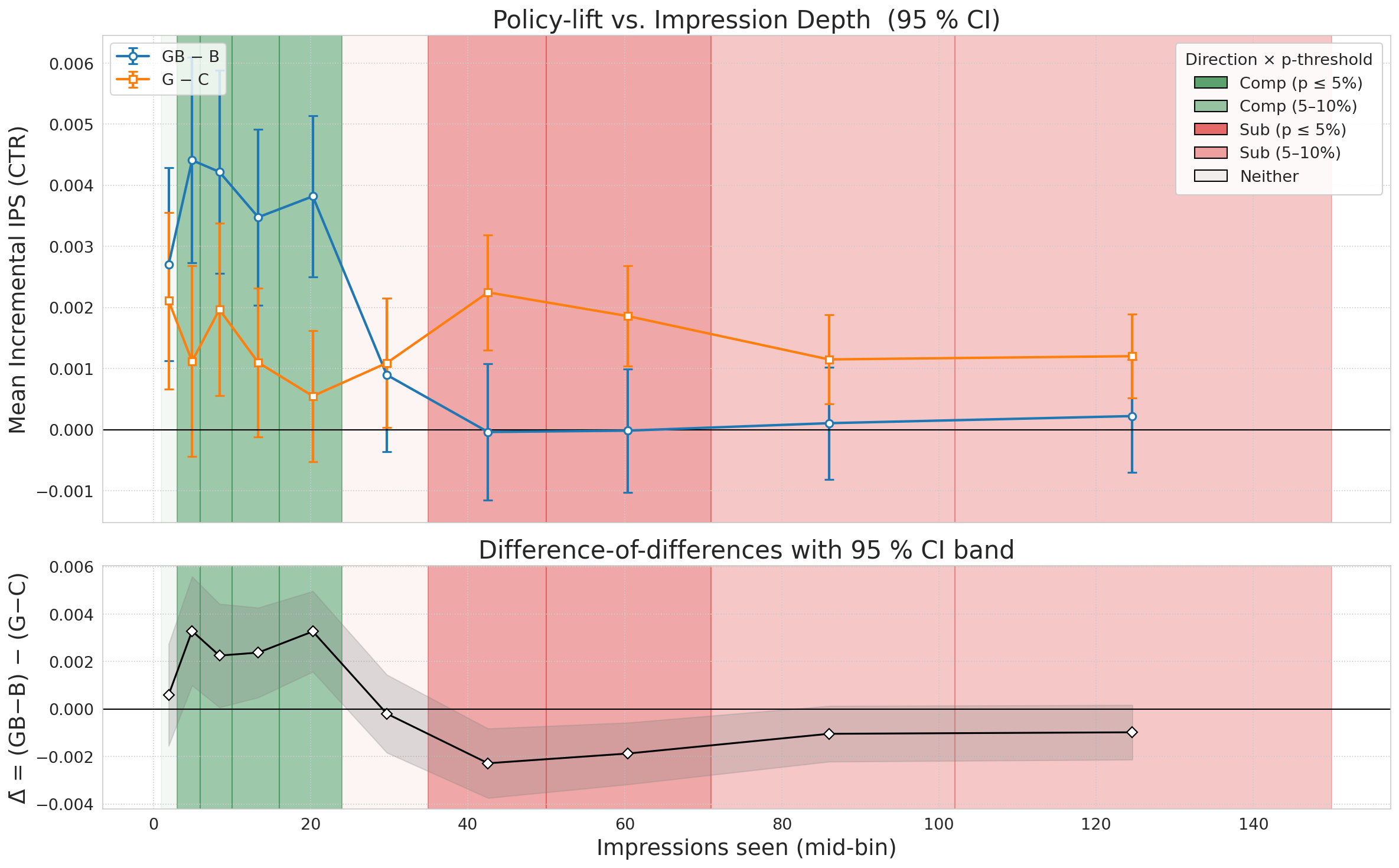}
    \caption{Complementarity and substitutability by impression depth. Top: $X^{GB} - X^B$ (blue) and $X^{G} - X^{\varnothing}$ (orange) with 95\% CIs. Bottom: difference-in-differences $\bar{\Delta}$ with 95\% CI; positive indicates complementarity, negative indicates substitutability.}
    \label{fig:comsub_depth}
\end{figure}

Building on the aggregate results in Table~\ref{tab:aggregate_comp_sub}, we now examine how the relationship between $X^B$ and $X^G$ varies with the number of impressions a user has seen. We group impressions into bins of equal size, each containing the same number of observations, defined as the number of prior ads shown to the same user. For each bin, we compute two policy-value differences: (i) the value difference between the $X^{GB}$ and $X^B$ policies, and (ii) the value difference between the $X^{G}$ and $X^{\varnothing}$ policies.

Figure~\ref{fig:comsub_depth} summarizes these patterns. The top panel plots both value differences across impression depth with 95\% confidence intervals, while the bottom panel reports their difference-in-differences, \(
\bar{\Delta} = (X^{GB} - X^B) - (X^{G} - X^{\varnothing}),\) where positive values indicate complementarity (joint value exceeds additivity) and negative values indicate substitutability (information sets overlap in value). Shaded regions denote ranges with statistically significant effects.

The figure reveals three stages. In the \emph{minimal behavioral history} stage (approximately 1--2 impressions), \(X^B\) contains little behavioral information. As a result, the stand-alone contribution of geographical data \(X^{G} - X^{\varnothing}\) is at its highest. In this range, adding geographical data to behavioral data delivers nearly the same gain as geography alone, so \(X^{GB} - X^{B} \approx X^{G} - X^{\varnothing}\). Consequently, \(X^{GB}\) yields little additional value relative to \(X^{G}\).

In the \emph{sparse behavioral history} stage (up to roughly 25 impressions), geographical information continues to play an important role as \(X^B\) begins to capture meaningful signals that, when combined with \(X^G\), create synergy between the two information sources. In this range, geographical and behavioral information act as complements: the value difference \(X^{GB} - X^B\) exceeds \(X^{G} - X^{\varnothing}\), which indicates that adding \(X^G\) to \(X^B\) delivers incremental targeting value beyond their separate contributions.

Finally, in the \emph{rich behavioral history} stage (from the low 20s and beyond), the marginal contribution of geographical data declines as accumulated behavioral information becomes substantially more informative. In this regime, the marginal contribution of adding \(X^G\) to \(X^B\) turns negative, indicating substitutability, as \(X^B\) already captures much of the variation that \(X^G\) would otherwise provide. For the full set of statistical test results underlying this analysis, we refer readers to Web Appendix~\S\ref{app:ComSubTests}.

\section{Mechanisms Underlying the Role of Geographical Data}\label{sec:geo_mechanismm}
%We now turn to the mechanisms that explain why geographical data ($X^G$) contributes to targeting outcomes in relation to behavioral data ($X^B$). Our unified framework has shown how the two information sets act as complements or substitutes in the decision value they generate. What remains unresolved is \emph{why} geographical data improves performance in certain cases: does $X^G$ capture an independent channel of information about user responsiveness, or does it primarily serve as a proxy for preference patterns that behavioral data eventually reveal? To address this question, we investigate the channels through which a user’s location can shape the probability of clicking on an ad. Prior work in marketing and social networks highlights three broad sources of spatially correlated responses: homophily, localized influence, and contextual confounding \citep[e.g.,][]{lovett2013brands, aral2012identifying}. Building on this classification, we distinguish three channels through which geographical data can affect ad responsiveness and consider, for each, the extent to which behavioral data may eventually capture the same information:

We now turn to the mechanisms that explain why geographical data ($X^G$) contributes to targeting outcomes in relation to behavioral data ($X^B$). Our unified framework has shown how the two information sets act as complements or substitutes in the decision value they generate. What remains unresolved is \emph{why} geographical data improves performance in certain cases: does $X^G$ capture an independent channel of information about user responsiveness, or does it primarily serve as a proxy for preference patterns that behavioral data eventually reveal?

To address this, we decompose spatial correlation in ad responsiveness into two distinct sources. The first is \emph{influence}, which has a causal interpretation: users respond similarly \emph{because} they are in the same location, which enables peer interaction and local information transmission. The second is \emph{confounding}: users respond similarly \emph{not because} of the same location itself, but because location is confounded  with other factors, such as similar characteristics, income, demographics, or cultural norms, that independently shape preferences. This distinction mirrors the social network literature, which distinguishes \emph{social influence} from \emph{latent homophily} as alternative explanations for correlated behavior among connected individuals \citep{anagnostopoulos2008influence, ma2015latent}. We present the two sources of existing spatial correlation as follows:

\begin{enumerate}

\item \textit{Spatial influence.} Under the causal interpretation, users may affect the decisions of others around them. For example, a user who sees an in-app advertisement for a product or promotion may mention it to friends or coworkers nearby, increasing their likelihood of engaging with the same ad when they encounter it. In this case, correlated responses arise from information spreading through local interactions rather than from shared preferences. Detecting such effects requires observing the spatial network that connects individuals. Geographical data ($X^G$) provides this information by revealing which users are located near one another and therefore more likely to interact. Behavioral data ($X^B$), which records only each user's actions, cannot capture these local interaction patterns. Consequently, if spatial influence exists, it can only be detected through geographical data that allows the underlying spatial network to be constructed.

\item \textit{Spatial confounding.} Under the confounding interpretation, users respond similarly not because they influence one another, but because location is correlated with shared characteristics that shape preferences. For example, residents of a wealthy neighborhood may be more likely to click on luxury-product ads because they have similar income levels and consumption preferences. Geographical data ($X^G$) captures this pattern immediately, since location acts as a proxy for the bundle of characteristics associated with that area. However, these preferences are gradually revealed through individual behavioral data ($X^B$). A user who repeatedly engages with premium brands or high-end retailers reveals the same underlying preference that geography initially proxies. As behavioral histories accumulate, models can infer these preferences directly from past actions, making geographical data primarily an early but coarse proxy whose value declines as behavioral data becomes richer.

\end{enumerate}

%In both cases, geographical data improves targeting when ad responses exhibit spatially structured variation that is not yet explained by behavioral data. To formalize this, we express click propensity as:
%\begin{equation}\label{equ:Spatial}
%y = f(X^B) + \varepsilon,
%\end{equation}
%where $f(X^B)$ represents the component of $y$ explained by behavioral data, and $\varepsilon$ captures the residual variation unexplained by $X^B$. Geographical data $X^G$ is informative precisely when it explains systematic spatial structure in $\varepsilon$, whether that structure arises from spatial homophily or influence. The key question, therefore, is whether this residual spatial structure persists as behavioral histories become richer. If geographical data primarily proxies for spatial homophily, then richer $X^B$ should absorb the clustering and leave little residual role for $X^G$. If, instead, geographical data contain spatial influence, spatial correlation should remain even after controlling for behavioral data. In what follows, we evaluate this distinction directly. \mohammad{re-writtn}

%In both cases, geographical data improves targeting when ad responses exhibit spatially structured variation not yet explained by behavioral data. 

Confounding in location intuitively means that the location selection by individuals is a function of underlying characteristics that are also correlated with the outcome of interest. 
To formalize this, recall $\omega \in \Omega$ that denote the true latent type of a user defined in \S\ref{sec:ProblemDef}. The true click propensity is therefore a function of this latent type, which we denote by $g(\omega)$. Since $\omega$ is unobserved, the platform relies on observable behavioral data $X^B$ to estimate this function. Let $f(X^B)$ denote the platform's estimate of $g(\omega)$ based on the available behavioral history. We can therefore write click propensity as
\begin{equation}\label{equ:Spatial}
y = g(\omega) + \theta_I + \varepsilon   = f(X^B) + \underbrace{\big[g(\omega) - f(X^B)\big]}_{\theta_C :~\text{estimation error}} + \theta_I + \varepsilon
\end{equation}
where $\varepsilon$ captures idiosyncratic variation in responses. In Equation~(\ref{equ:Spatial}), spatial influence is captured by $\theta_I$, while spatial confounding appears in the estimation error $\theta_C = g(\omega)-f(X^B)$. Geographical data $X^G$ is informative when spatial correlation remains in these terms after conditioning on behavioral data $X^B$. Under spatial influence, users affect the responses of nearby users through local interactions. These effects cannot be disentangled from individual preferences, regardless of how rich $X^B$ becomes. If influence exists, it creates spatial correlation in $\theta_I$ that $X^G$ can capture. Under spatial confounding, location is correlated with latent characteristics in $\omega$ that affect ad responsiveness. When behavioral histories are limited, $f(X^B)$ does not fully capture these characteristics, leaving spatial variation in $\theta_C=g(\omega)-f(X^B)$ that $X^G$ can explain. As behavioral histories accumulate and $f(X^B)$ better approximates $g(\omega)$, this component shrinks. The key empirical question is whether the predictive value of $X^G$ disappears as behavioral histories become richer or persists after conditioning on $X^B$.

\subsection{Empirical Strategy: Residualized Spatial Autocorrelation Test}\label{ssec:geo_empirical}
To evaluate this decomposition, we test whether spatial correlation remains after conditioning on behavioral data $X^B$. In Equation~(\ref{equ:Spatial}), spatial confounding enters through $\theta_C = g(\omega)-f(X^B)$ and should diminish as behavioral histories allow $f(X^B)$ to better approximate $g(\omega)$. Spatial influence, captured by $\theta_I$, instead arises from local interactions and therefore persists regardless of how rich $X^B$ becomes. Our empirical strategy therefore tests whether residual spatial correlation disappears once behavioral information is incorporated. If it does, geographical data primarily reflects spatial confounding already absorbed by $X^B$; if it persists, it indicates spatial influence. We implement this idea using a residualized spatial autocorrelation (RSA) test.

Our empirical strategy proceeds by aggregating outcomes at the regional level. Let $Y_c$ denote the total number of clicks and $I_c$ the total number of impressions in region $c$. Each impression can be viewed as a Bernoulli trial with region-specific click probability $p_c$, but because click events are rare, the Binomial distribution is well approximated by a Poisson model:
\[
Y_c \mid p_c \sim \text{Poisson}(I_c \, p_c).
\]
This specification is standard in spatial count-data analysis. The offset term $\log(I_c)$ adjusts for heterogeneous exposure across regions, while the residuals from the fitted model provide a natural basis for testing whether unexplained variation in click-through rates is spatially correlated.

\squishlist
    \item \textbf{Step 1: Baseline spatial structure in raw CTR:} As a starting point, we examine whether CTRs exhibit geographical clustering in the absence of any behavioral controls. We estimate a Poisson count model that includes only an exposure offset and define the residual:
    \[
    Y_c \sim \text{Poisson}(\mu_c), \quad \log \mu_c = \alpha + \log(I_c), \quad 
    {\varepsilon^{(0)}_c = \log(Y_c/I_c) - \alpha},
    \]
    Here, $\log(I_c)$ ensures that $\mu_c / I_c$ corresponds to the expected CTR in region $c$, while $\varepsilon^{(0)}_c$ measures deviations from the global mean CTR after adjusting for exposure volume. Evidence of spatial autocorrelation in $\varepsilon^{(0)}_c$ would indicate that raw CTRs contain location-based structure, consistent with spatial confounding or influence.

    \item \textbf{Step 2: Residual spatial structure after controlling for behavior:} Next, we ask whether this spatial structure remains once behavioral data are taken into account. To do so, we augment the model with the fitted behavioral component $f(X^B_c)$ and compute the adjusted residual:
    \[
    Y_c \sim \text{Poisson}(\mu_c), \quad \log \mu_c = \alpha + f(X^B_c) + \log(I_c), \quad 
    {\varepsilon^{(B)}_c = \log(Y_c/I_c) - \alpha - f(X^B_c)},
    \]
    This residual captures variation in CTR unexplained by either exposure or behavioral features. If $f(X^B_c)$ successfully absorbs location-linked patterns, spatial dependence in $\varepsilon^{(B)}_c$ should be weaker than in $\varepsilon^{(0)}_c$.

    \item \textbf{Step 3: Testing for spatial dependence:} Finally, we test whether $\varepsilon^{(0)}_c$ and $\varepsilon^{(B)}_c$ exhibit spatial autocorrelation. We apply two standard measures: Moran’s $I$ \citep{moran1950notes}, which captures global correlation across regions, and Geary’s $C$ \citep{geary1954contiguity}, which is more sensitive to local clustering. For a generic set of residuals $\varepsilon_c$ indexed by region $c$, these are defined as:
    \[
    I = \frac{N}{\sum_{i} \sum_{j} w_{ij}} \cdot 
    \frac{\sum_{i} \sum_{j} w_{ij} (\varepsilon_i - \bar{\varepsilon})(\varepsilon_j - \bar{\varepsilon})}
    {\sum_{i} (\varepsilon_i - \bar{\varepsilon})^2}, 
    \quad 
    C = \frac{(N - 1) \sum_{i} \sum_{j} w_{ij} (\varepsilon_i - \varepsilon_j)^2}
    {2 \sum_{i} (\varepsilon_i - \bar{\varepsilon})^2},
    \]
    where $N$ is the number of regions, $w_{ij}$ denotes the $(i,j)$ element of the spatial weight matrix $W$, and $\bar{\varepsilon}$ is the mean of the residuals. Significant positive values of Moran’s $I$ (or values of Geary’s $C$ below one) indicate that geographically proximate regions have similar residuals. By comparing the statistics for $\varepsilon^{(0)}_c$ and $\varepsilon^{(B)}_c$, we can assess whether behavioral features reduce spatial dependence, thereby clarifying whether geographical data provides independent information beyond what behavior explains.

\squishend

\subsection{Results: Spatial Correlation in CTRs}\label{ssec:geo_results}
We present two sets of results. First, in \S\ref{sssec:geo_baseline}, we examine spatial correlation in baseline versus behavior-adjusted CTR residuals. Second, in \S\ref{sssec:geo_sparse_rich}, we compare spatial dependence under sparse versus rich behavioral histories. Together, these analyses show how the role of geographical data changes once behavioral information is taken into account.

\subsubsection{Residual Spatial Dependence in Baseline and Behavior-Adjusted Models}\label{sssec:geo_baseline}
We begin by applying the RSA framework introduced in \S\ref{ssec:geo_empirical}. Specifically, we compare spatial correlation in the baseline residuals \(\varepsilon^{(0)}_c\), obtained without behavioral controls, to the behavior-adjusted residuals \(\varepsilon^{(B)}_c\). This analysis is conducted at both the county and city levels to assess how much of the observed spatial dependence can be explained by behavioral data.

\begin{figure}[htp!]
    \centering
    \includegraphics[width=0.9\linewidth]{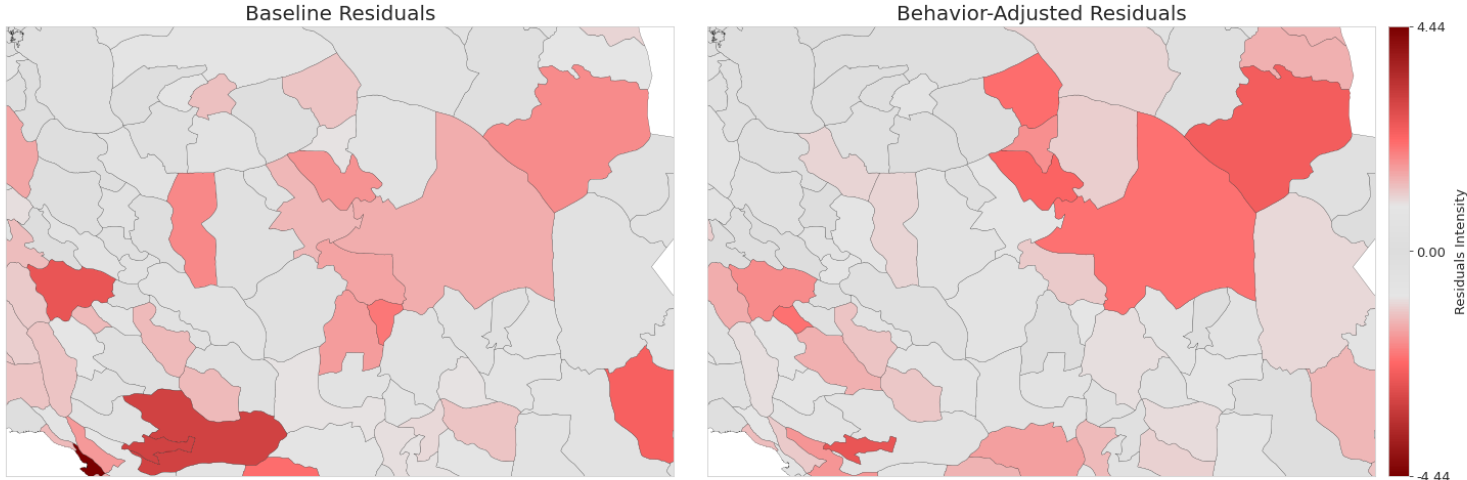}
    \caption{Spatial distribution of baseline (left) and behavior-adjusted (right) residuals at the county level. Darker red shades indicate higher residual intensity. Behavioral adjustment visibly reduces, but does not fully eliminate, spatial clustering.}
    \label{fig:residual_maps}
\end{figure}

Figure~\ref{fig:residual_maps} provides an intuitive visualization for the county-level case. The left panel displays the baseline residuals $\varepsilon^{(0)}_c$ from the model without behavioral controls. Several contiguous regions show clusters of similarly high or low residual values, indicating clear spatial autocorrelation. Such clustering is consistent with the presence of spatial homophily and/or influence. In contrast, the right panel shows the behavior-adjusted residuals $\varepsilon^{(B)}_c$. Incorporating $f(X^B_c)$ visibly reduces the strength and extent of spatial clusters, particularly in the southwest, although notable pockets remain. This pattern suggests that behavioral data absorb a substantial share of the location-linked variation, but not all of it.

To quantify the patterns observed in Figure~\ref{fig:residual_maps}, Table~\ref{tab:spatial_corr} reports Moran’s $I$ and Geary’s $C$ at both county and city levels. At the county level, the baseline residuals $\varepsilon^{(0)}_c$ exhibit strong and significant spatial dependence: regions with unusually high (or low) CTR residuals tend to be geographically proximate. After controlling for behavioral information, both statistics decline markedly, indicating that much of this clustering is explained by differences in user behavior. However, the remaining positive spatial correlation implies that geographical data continues to capture additional variation not yet absorbed by behavior.

At the city level, the attenuation is sharper. Baseline residuals still display detectable clustering, but once behavioral features are included, both Moran’s $I$ and Geary’s $C$ fall to levels that are statistically indistinguishable from zero. This result suggests that at finer geographical resolution, most of the spatial variation in CTR can be accounted for by behavioral histories, leaving little independent role for geographical data. The contrast between county and city levels thus highlights how geographical data's incremental contribution diminishes as spatial units become more granular, consistent with the idea that geographical data largely serves as a proxy for spatial confounding that behavioral data can eventually capture.

\begin{table}[h]
\centering
\footnotesize{
\begin{tabular}{l|c|c|c|c}
 & \textbf{County-level, \(\varepsilon^{(0)}_c\)} & \textbf{County-level, \(\varepsilon^{(B)}_c\)} & \textbf{City-level, \(\varepsilon^{(0)}_c\)} & \textbf{City-level, \(\varepsilon^{(B)}_c\)} \\ \hline
Moran's $I$   & 0.122 & 0.084 & 0.055 & -0.001 \\
$p$-value     & 0.00006 & 0.00318 & 0.00242 & 0.49156 \\ 
Geary's $C$   & 0.891 & 0.900 & 0.948 & 0.996 \\
$p$-value     & 0.00094 & 0.00162 & 0.01646 & 0.41863
\end{tabular}
\caption{Moran's $I$ and Geary's $C$ statistics for spatial autocorrelation of residuals before (baseline) and after (behavior-adjusted) controlling for behavioral features, at county and city levels.} 
\label{tab:spatial_corr}}
\end{table}

Taken together, these results show that geographical data explains substantial spatial variation in ad responses when behavioral information is excluded, but much of this variation disappears once behavioral data are incorporated. Moreover, the remaining spatial structure becomes smaller at finer spatial resolutions. To examine whether this attenuation also depends on the amount of behavioral information available for each user, we next split impressions into sparse and rich behavioral-history subsets and repeat the analysis.

\subsubsection{Residual Spatial Dependence under Sparse versus Rich Behavioral Histories}\label{sssec:geo_sparse_rich}
As discussed earlier, the ability of the behavioral model $f(X^B)$ to account for variation in CTR depends on the amount of behavioral history available for each user. To examine this effect, we split the sample into two groups: (i) a \emph{sparse-history} group, consisting of the first 50\% of impressions observed for each user, and (ii) a \emph{rich-history} group, consisting of an equal number of later impressions for the same users. This design allows us to compare scenarios where targeting models rely on limited versus extensive behavioral histories.

\begin{figure}[htp!]
    \centering
    \includegraphics[width=0.9\linewidth]{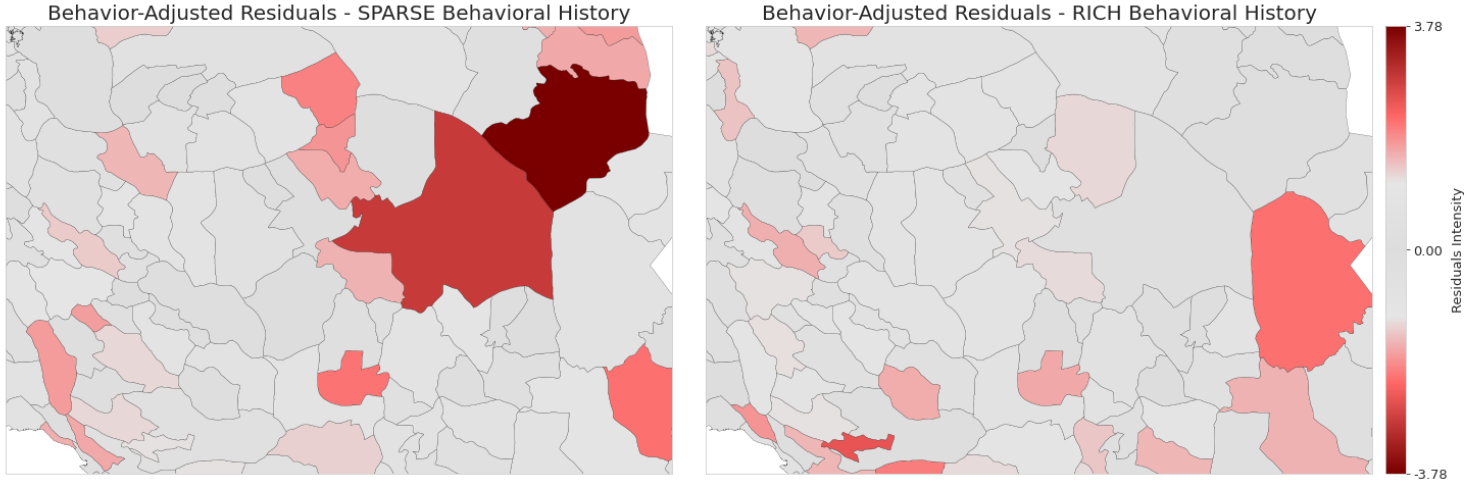}
    \caption{Behavior-adjusted residuals at the county level for sparse (left) and rich (right) behavioral histories. Darker shades indicate higher residual intensity.}
    \label{fig:sparse_rich_maps}
\end{figure}

Figure~\ref{fig:sparse_rich_maps} illustrates the spatial distribution of behavior-adjusted residuals at the county level for the two groups. In the sparse-history case (left panel), several large contiguous clusters of high residuals remain even after controlling for $f(X^B)$, indicating that location continues to explain a meaningful share of CTR variation. By contrast, in the rich-history case (right panel), residual clustering is visibly weaker and more fragmented, suggesting that much of the spatial dependence disappears once users accumulate longer behavioral records.

\begin{table}[h]
\centering
\footnotesize{
\begin{tabular}{l|c|c|c|c}
 & \textbf{County-level, Sparse} & \textbf{County-level, Rich} & \textbf{City-level, Sparse} & \textbf{City-level, Rich} \\ \hline
Moran's $I$   & 0.069 & 0.025 & -0.005 & 0.0005 \\
$p$-value     & 0.011 & 0.183 & 0.412 & 0.459 \\ 
Geary's $C$   & 0.952 & 0.976 & 1.012 & 1.006 \\
$p$-value     & 0.072 & 0.225 & 0.285 & 0.403
\end{tabular}
\caption{Moran's $I$ and Geary's $C$ statistics for spatial autocorrelation of behavior-adjusted residuals under sparse and rich behavioral histories, at county and city levels.} 
\label{tab:spatial_corr_rich_sparse}}
\end{table}

Table~\ref{tab:spatial_corr_rich_sparse} formalizes these visual patterns. Under sparse histories, Moran’s $I$ is positive and statistically significant at the county level, while Geary’s $C$ also indicates non-random clustering. These results confirm that when behavior is limited, geographical data captures residual variation that remains spatially structured. In the rich-history case, however, both statistics decline substantially in magnitude and lose statistical significance across county and city levels. This indicates that once behavioral data are rich, $f(X^B)$ already accounts for nearly all of the systematic spatial correlation in CTR. 

The comparison between sparse and rich behavioral histories also sheds light on the mechanisms through which geographical data matters. When behavioral histories are sparse, residual spatial structure may arise from spatial influence, spatial confounding, or both. Observing how this structure evolves as behavioral data becomes richer is therefore informative about which mechanism dominates. Spatial influence is inherently relational and cannot be recovered from individual behavioral data $X^B$, regardless of how rich those histories become. If spatial influence were the primary driver, residual spatial autocorrelation would persist even after conditioning on $X^B$. Instead, we find that spatial autocorrelation largely disappears once behavioral histories are rich. This pattern rules out spatial influence and suggests that spatial confounding is the dominant channel: geographical data initially captures similarities in preferences across nearby users, but these similarities are eventually learned directly from individual behavioral histories, leaving little residual role for $X^G$.

\section{Managerial and Policy Implications}\label{sec:Managerial}

From the perspective of advertising platforms and advertisers, our results highlight a clear trade-off in the use of geographical information for engagement modeling. Geographical data provides incremental value primarily during early cold-start phases, when behavioral histories are sparse and cannot yet support effective personalization. As behavioral data accumulates, it increasingly substitutes for the value previously provided by geographical information, making location-based targeting largely redundant. This pattern suggests that firms can restrict the use of geographical data to early cold-start stages and gradually transition to behavioral targeting as behavioral histories mature. Such an approach allows firms to reduce unnecessary collection of location information while maintaining targeting performance, and may also strengthen consumer trust through transparent, user-controlled opt-out mechanisms for geographical tracking \citep{martin2017data}.

From a consumer perspective, the substitutability of geographical information by behavioral information implies that consumers can safely opt out of location tracking once sufficient behavioral history has accumulated. In this regime, relevant targeting can be maintained without continued use of geographical information, allowing consumers to limit exposure of sensitive location attributes without materially affecting engagement outcomes.

From a regulatory perspective, the limited incremental value of geographical information in the presence of rich behavioral histories raises questions about the justification for continued large-scale location tracking. While geographical information may serve a temporary role in early cold-start settings, its long-run substitutability suggests that stricter oversight of persistent geographical tracking is warranted, particularly in digital advertising environments where comparable targeting performance can be achieved with behavioral information alone.

\section{Conclusion}\label{sec:conclusion}
In this paper, we develop a unified framework to evaluate the value of information in digital advertising, integrating economic theory, machine learning, and causal inference. Conceptually, we extend decision-based definitions of information value to test whether behavioral and geographical data act as complements or substitutes. Methodologically, we combine LSTM architectures with attention mechanisms to capture temporal and spatial dynamics and use inverse propensity scoring to recover counterfactual performance from observational data. Together, these elements provide a generalizable approach for comparing multiple information sets in high-dimensional environments, with applications extending beyond advertising.

%Our results show that both geographical and behavioral data enhance targeting, but their roles change as behavioral histories become richer. At the aggregate level, complementarities and substitutions offset one another, yielding no clear overall interaction. As users accumulate impressions, geographical data explains most variation when behavioral data contain almost no information, complements behavior when histories are sparse, and becomes a substitute once histories are rich. This progression indicates that the value of geographical data is temporary, concentrated in the early stages before behavioral information alone suffices for accurate targeting.

Our results show that both geographical and behavioral data enhance targeting, but their roles change as behavioral histories become richer. At the aggregate level, complementarities and substitutions offset one another, yielding no clear overall interaction. Examining heterogeneity as users accumulate behavioral histories allows us to uncover a clear pattern. Geographical data explains most variation when behavioral information is minimal, complements behavioral data when histories are sparse, and becomes a substitute once histories are rich. This progression indicates that the value of geographical data is temporary and concentrated in the early stages before behavioral information alone suffices for accurate targeting.

%To explore the underlying mechanism, we develop a Residualized Spatial Autocorrelation test. We first document strong spatial clustering in ad responses, consistent with nearby users exhibiting similar preferences. Once outcomes are residualized on behavioral data, much of this correlation vanishes, indicating that repeated user actions absorb the underlying similarity. Comparing sparse and rich histories, residual clustering persists only when behavioral information is limited but disappears once histories are extensive. These results suggest that the apparent value of geography is primarily driven by homophily that behavioral data ultimately capture. We do not find evidence of an independent role for geography beyond this channel and make no claims regarding other mechanisms such as influence or local shocks.

Future research could extend our analysis in several ways. First, beyond in-app advertising, the value of geographical data may differ in sectors where location remains central, such as transportation, local retail, or urban planning. Second, future work should trace the privacy–utility frontier by evaluating policy value under explicit data-use constraints, including coarse geolocation, $k$-anonymity, differential privacy, or federated learning. Finally, our analysis focuses on online in-app advertising, where engagement is primarily shaped by behavioral information rather than physical location. We do not generalize these conclusions to industries in which location carries intrinsic value, such as ride-hailing, food delivery, local event promotions, or brick-and-mortar retail discounts. In such contexts, geographical information may remain a critical input, and further research is needed to assess whether behavioral information can fully substitute for location-based targeting.

%These extensions would deepen insights into how visual media influences public discourse and societal outcomes.

\section*{Competing Interests Declaration}
Author(s) have no competing interests to declare.

%\singlespacing
%    \setlength{\bibsep}{0pt plus 5ex}
%\bibliographystyle{plainnat}
%\bibliography{sample}

%\input{sample.bbl}  % Use your actual .bbl file name

{\singlespacing

}

\setcounter{table}{0}
\setcounter{page}{0}
\setcounter{figure}{0}

\renewcommand{\thetable}{A\arabic{table}}
\renewcommand{\theequation}{A.\arabic{equation}}
\renewcommand{\thepage}{\roman{page}}
\renewcommand{\thefigure}{A.\arabic{figure}}

\renewcommand{\thepage}{\roman{page}}
\newpage
\renewcommand*\appendixpagename{Web Appendix}

\begin{appendices}

\section{Details of Machine Learning Framework}\label{app:ML} 
In this section, we model \( f_X \) by specifying both the information set \( X \) and the functional form \( f(\cdot) \). We discuss the specific features included in each informational regime, \( \{X^\varnothing, X^G, X^B, X^{GB}\} \), in \S\ref{sssec:X}. We then describe the implementation details of behavioral modeling using an LSTM with an attention mechanism, with additional details provided in \S\ref{app:LSTM}.

\subsection{Structure of Information Sets $X$}\label{sssec:X}
For our machine learning model, each targeting regime $X^{(\cdot)}$ corresponds to a specific set of observable features. We classify these features into three dimensions, \textit{contextual}, \textit{behavioral}, and \textit{geographical}, which form the building blocks of the different regimes:

\textit{Contextual Features (\( X^C \)):}
Contextual features characterize the \emph{setting} in which each ad impression occurs, independent of user identity or location. These include time of day, day of week, app, device, and network provider. Engagement often follows systematic patterns along these dimensions; for instance, click rates in gaming apps may peak in the evening when many users log in.

\textit{Behavioral Features (\( X^{\text{Behave}} \)):}
Behavioral features capture a user's \emph{online footprint}, such as prior ad exposures, cumulative clicks, time since last click, and category-specific CTRs. These variables reveal individual preferences and engagement; for example, a user who frequently clicks on sports ads is much more likely to respond to a new sports campaign.

\textit{Geographical Features (\( X^{\text{Geo}} \)):}
Geographical features reveal the user’s \emph{physical footprint}, their location at the time of the impression, which may be represented at various levels of granularity, such as city or precise latitude and longitude. They reflect spatial variation in preferences and engagement, for example, ride-sharing ads see higher click rates in dense urban centers during rush hour compared to rural areas.

While we have conceptually defined the targeting regimes, here we map them to their feature representations. Each \( X^{(\cdot)} \) corresponds to a specific subset of contextual, behavioral, and geographical features used in estimation:

\begin{itemize}
    \item \textbf{(\( X^\varnothing = X^C \)):} 
    The advertiser observes only contextual features (e.g., device, app category, time, or network conditions). No user-level information, such as behavioral or geographical data, is available, so actions are chosen based on context-dependent averages rather than latent user types~$\omega$. Such information, typically available from campaign reporting or platform analytics.

    \item \textbf{(\( X^B = X^{\text{Behave}} \cup X^C \)):} The advertiser observes both the context of each impression and features that summarize the user’s prior behavior. Together, these variables provide insight into the user's likely preferences and engagement, revealing information about the latent type \(\omega\) through observed patterns.

    \item \textbf{(\( X^G = X^{\text{Geo}} \cup X^C \)):} The advertiser observes both contextual information and the user’s geographical location. Geographical features can reveal spatial heterogeneity in preferences, such as differences due to local events, regional trends, or time zones, and thus offer indirect information about a user’s type \(\omega\).

    \item \textbf{(\( X^{GB} = X^{\text{Behave}} \cup X^{\text{Geo}} \cup X^C \)):} The advertiser has access to the full set of behavioral, geographic, and contextual features. This regime enables the model to capture the most complete picture of user heterogeneity to infer latent user types  \(\omega\) and guide targeting.
\end{itemize}

These categories mirror the types of data advertisers realistically access in practice. The no-information case serves as a benchmark for non-personalized targeting, while the other regimes reveal distinct dimensions of latent user preferences. Details on feature construction, variable definitions, and preprocessing appear in the Web Appendix (\S\ref{app:Features}).

\subsection{Behavioral Learning Algorithm Implementation: LSTM}\label{app:LSTM}
To effectively model the sequential and temporal dynamics of user engagement, we employ the Long Short-Term Memory (LSTM) predictive model, a specialized class of Recurrent Neural Networks (RNNs) \citep{rumelhart1986learning}. LSTMs, first introduced by \citet{hochreiter1997long}, address the limitations of traditional RNNs, such as the vanishing gradient problem, by incorporating a memory cell and gating mechanisms that regulate the flow of information. These features make LSTMs highly effective for tasks involving long-term dependencies and sequential patterns. LSTMs have been widely adopted in user engagement modeling due to their ability to handle sequential dependencies and dynamic user behavior \citep{grbovic2018real, quadrana2018sequence}.

An LSTM unit maintains two key components at each time step \( t \): the cell state \( C_t \), which acts as a long-term memory reservoir, and the hidden state \( h_t \), which encapsulates information relevant for the current time step. The evolution of these components is controlled by three gating mechanisms: the forget gate, the input gate, and the output gate, as defined:

\begin{itemize}
    \item \textbf{Forget Gate.} The forget gate determines how much of the previous cell state \( C_{t-1} \) should be retained:
    \[
    f_t = \sigma(W_f [h_{t-1}, x_t] + b_f),
    \]
    where \( f_t \) is the forget gate vector, \( h_{t-1} \) is the hidden state from the previous time step, \( x_t \) is the input at the current time step, \( W_f \) and \( b_f \) are learnable parameters (weights and biases), and \( \sigma(\cdot) \) is the sigmoid activation function, mapping values to the range \([0, 1]\).

    \item \textbf{Input Gate.} The input gate determines the extent of new information to be incorporated into the memory cell. It is defined as:
    \[
    i_t = \sigma(W_i [h_{t-1}, x_t] + b_i),
    \]
    \[
    \tilde{C}_t = \tanh(W_C [h_{t-1}, x_t] + b_C),
    \]
    where \( i_t \) is the input gate vector, \( \tilde{C}_t \) is the candidate cell state, and \( W_i, W_C, b_i, b_C \) are trainable parameters. The hyperbolic tangent (\( \tanh \)) maps the candidate state values to \([-1, 1]\).

    \item \textbf{Cell State Update.} The cell state \( C_t \) is updated by combining the contributions of the forget and input gates:
    \[
    C_t = f_t \odot C_{t-1} + i_t \odot \tilde{C}_t,
    \]
    where \( \odot \) denotes element-wise multiplication.

    \item \textbf{Output Gate.} The output gate determines the current hidden state \( h_t \), which is used to generate the output and influence subsequent computations:
    \[
    o_t = \sigma(W_o [h_{t-1}, x_t] + b_o),
    \]
    \[
    h_t = o_t \odot \tanh(C_t),
    \]
    where \( o_t \) is the output gate vector and \( W_o, b_o \) are learnable parameters.
\end{itemize}

These equations collectively enable the LSTM to selectively retain, update, and output information across time steps, making it robust to long-term dependencies in sequential data. 

In addition to the recurrent backbone, our predictive framework integrates several complementary components, each designed to address a specific challenge in sequential engagement modeling. First, \textit{embedding layers} transform high-cardinality categorical inputs, such as device type, province, city, or network provider, into low-dimensional dense vectors. This approach not only reduces dimensionality but also enables the model to learn latent similarity structures among categories, improving generalization and mitigating the curse of dimensionality \citep{guo2016entity}. Second, temporal structure is enriched through two mechanisms: (i) \textit{absolute positional embeddings}, which provide the model with awareness of an event’s position within the sequence, and (ii) \textit{time–gap embeddings}, which capture irregular intervals between events, allowing differentiation between dense bursts of activity and prolonged inactivity. Third, a \textit{stacked LSTM} serves as the primary sequence encoder, effectively modeling order-sensitive dependencies and evolving behavioral patterns over time. Fourth, a \textit{causal multi-head self-attention} layer enables the model to focus adaptively on past events that are most relevant to the current prediction, capturing long-range dependencies while enforcing strict causality constraints \citep{vaswani2017attention}. Finally, a \textit{gated projection head} selectively filters and transforms the attended sequence representations before a linear readout layer maps them to click probabilities, enhancing the interpretability and robustness of the final decision stage. Table~\ref{tab:nas_sequence_model_architecture} reports the architecture and parameter counts.

\begin{table}[htp!]
\centering
\caption{Summary of the Sequence Model Architecture and Parameters}
\begin{tabular}{llr}
\hline
\textbf{Layer (type)} & \textbf{Output Shape} & \textbf{Param \#} \\ \hline
Input Features & [-1, 150, 240] & 0 \\
Time-Gap Projection & [-1, 150, 240] & 57,840 \\
Absolute Positional Embedding & [1, 150, 240] & 36,000 \\
LSTM (4 layers) & [-1, 150, 512] & 4,269,568 \\
Layer Norm (pre-attention) & [-1, 150, 512] & 1,024 \\
Causal Multi-Head Attention & [-1, 150, 512] & 1,049,600 \\
Layer Norm (post-attention) & [-1, 150, 512] & 1,024 \\
Gate Projection (sigmoid) & [-1, 150, 512] & 262,656 \\
Body Projection (tanh) & [-1, 150, 512] & 262,656 \\
Dropout & [-1, 150, 512] & 0 \\
Final FC Output Layer & [-1, 150, 1] & 513 \\ \hline
\textbf{Total Parameters} &  & \textbf{5,941,905} \\ \hline
\end{tabular}
\label{tab:nas_sequence_model_architecture}
\end{table} 
In the following, we provide a detailed, step-by-step description of the architecture, formally defining its components and explaining how they interact to transform raw sequential inputs into engagement probability forecasts.

We model user engagement as a sequential decision prediction problem, where the state at each time step incorporates both numerical and categorical information observed up to that point. Let \( j \) index users and \( t \in \{1, \dots, T\} \) denote the position in the sequence of impressions observed for user \( j \), with a maximum sequence length \( T = 150 \).

\paragraph{Input Representation.}  
At each time step \( t \), we observe:
\[
X^{\mathrm{num}}_{jt} \in \mathbb{R}^n, 
\quad \{ X^{\mathrm{cat}, k}_{jt} \}_{k=1}^c,
\]
where \( n \) is the number of numerical features and \( c \) is the number of categorical features. Numerical features capture continuous signals (e.g., time-based statistics, historical engagement rates, contextual metrics). Categorical features correspond to high-cardinality identifiers such as device model, province, city, or advertisement ID. Direct one-hot encoding of categorical features would be prohibitively high-dimensional and sparse. Instead, we map each categorical variable \( X^{\mathrm{cat}, k}_{jt} \) to a dense, learnable embedding vector:
\[
\mathbf{e}^{(k)}_{jt} \in \mathbb{R}^{d_k}, 
\]
where \( d_k \) is the embedding dimension for category \( k \). Embeddings enable the model to learn latent similarity structures, e.g., provinces with similar user behavior or devices with similar performance profiles, without manual feature engineering. Concatenating all categorical embeddings yields:
\[
\mathbf{e}_{jt} = [\mathbf{e}^{(1)}_{jt}, \dots, \mathbf{e}^{(c)}_{jt}] 
\in \mathbb{R}^{E_{\mathrm{cat}}}, 
\quad E_{\mathrm{cat}} = \sum_{k=1}^{c} d_k.
\]
The full feature vector at time \( t \) is:
\[
X_{jt} = [ X^{\mathrm{num}}_{jt}, \mathbf{e}_{jt} ] \in \mathbb{R}^{d_{\mathrm{in}}},
\quad d_{\mathrm{in}} = n + E_{\mathrm{cat}}.
\]
\paragraph{Temporal Augmentation.}  
To model sequence position and irregular arrival times, we augment \( X_{jt} \) with two temporal components:
\begin{enumerate}
    \item \textit{Absolute positional embedding} \( \mathbf{P}_t \in \mathbb{R}^{d_{\mathrm{in}}} \), a learnable vector for each position \( t \), allows the model to differentiate between early- and late-stage interactions even if other features are identical.
    
    While LSTMs can track order, they do so implicitly through hidden state transitions. Positional embeddings make temporal position an explicit feature, which is helpful in cases where position itself carries meaning (e.g., “first impressions” vs. “later impressions” behave differently). LSTM gates learn relative dependencies; positional embeddings give a direct absolute time index, which is crucial for long sequences where LSTM memory may fade.
    \item \textit{Time–gap embedding} models the elapsed time since the previous impression:
    \[
    \Delta t_{jt} = \text{time}(j,t) - \text{time}(j,t-1),
    \]
    transformed as \( \log(1 + \Delta t_{jt}) \) and projected to \( \mathbb{R}^{d_{\mathrm{in}}} \) through a learned linear layer. This captures behavioral intensity, distinguishing between dense bursts of activity and long idle periods.
    
    The spacing between events often changes behavior (e.g., if a user sees an ad after 5 seconds vs. after 5 days). Neither positional embeddings nor LSTM hidden states naturally capture this irregular spacing. Positional embeddings tell us “this is the 10th impression,” but not whether it came an hour or a week after the previous one. Time–gap embeddings fill that gap.
\end{enumerate}
The resulting time-aware token is:
\[
\tilde{X}_{jt} = X_{jt} + \mathbf{P}_t + \mathrm{TimeGapProj}\big( \log(1 + \Delta t_{jt}) \big).
\]

\paragraph{Sequence Encoding.}  
The augmented sequence \(\{\tilde{X}_{j1}, \dots, \tilde{X}_{jT}\}\) is processed by a stacked LSTM with hidden size \(H = 512\) and \(L_{\mathrm{LSTM}} = 4\) layers:
\[
h^{(\ell)}_{jt} = \mathrm{LSTM}_\ell\big(h^{(\ell-1)}_{jt}\big),
\]
where \(h^{(0)}_{jt} = \tilde{X}_{jt}\). The LSTM captures local and medium-range temporal dependencies, preserving order-sensitive dynamics such as gradual preference shifts or decaying effects of prior interactions. Attention alone doesn’t model sequential dependencies as naturally as LSTMs, and attention without recurrence can sometimes struggle with shorter-term patterns when data is noisy. The LSTM provides a strong temporal backbone before attention refines it.

\paragraph{Causal Multi-Head Attention.}  
To model longer-range dependencies and non-local feature interactions, we apply a causal multi-head self-attention layer with \(H_{\mathrm{attn}} = 4\) heads to the LSTM outputs. Attention reweights historical states based on their relevance to the current prediction. We enforce causality with a triangular mask:
\[
\mathrm{mask}(t',t) =
\begin{cases}
0, & t' \le t, \\
-\infty, & t' > t,
\end{cases}
\]
ensuring that predictions at time \(t\) use only past and present information. A key–padding mask derived from \(m_{jt} \in \{0,1\}\) ignores padded steps. Layer normalization before and after attention stabilizes optimization.

LSTMs compress history into a single hidden state, which can lose fine-grained details over long sequences. Attention re-opens the past and selectively retrieves important past events directly. It complements the LSTM by providing content-based memory access useful when long-range dependencies matter, e.g., a behavior 100 impressions ago is relevant now.

\paragraph{Gated Projection Head.}  
The attention-enhanced representation is passed through a gated projection mechanism, which adaptively filters and transforms features:
\[
g_{jt} = \sigma(W_g h_{jt}), \quad b_{jt} = \tanh(W_b h_{jt}),
\]
\[
z_{jt} = \mathrm{Dropout}(g_{jt} \odot b_{jt}).
\]
Here, the gate vector \(g_{jt}\) acts as a learnable feature selector, while \(b_{jt}\) generates candidate transformations. Their elementwise product modulates information flow, suppressing irrelevant history and emphasizing predictive patterns. 

It lets the model selectively emphasize or suppress certain dimensions in the representation before making predictions, essentially a learned attention mechanism at the feature level. Attention looks across time, but this gating looks within the feature vector, so it’s a different form of selectivity.

\paragraph{Prediction Layer.}  
The gated features are mapped to logits:
\[
\tilde{y}_{jt} = w^\top z_{jt}, \quad
\hat{y}_{jt} = \sigma(\tilde{y}_{jt}),
\]
where \(\hat{y}_{jt}\) is the predicted click probability at time \(t\).

\paragraph{Training Objective.}  
We minimize the binary cross-entropy loss with logits:
\[
\mathcal{L}(\Theta) = \sum_{j=1}^{n} \sum_{t=1}^{T} m_{jt} \,
\ell_{\mathrm{BCE\text{-}logits}}(y_{jt}, \tilde{y}_{jt}),
\]
where \(m_{jt}\) masks out padded steps, and \(\Theta\) includes all learnable parameters (embeddings, positional encodings, LSTM, attention, gating, and readout). Optimization uses AdamW \citep{Loshchilov2019Decoupled}, which combines adaptive learning rates with decoupled weight decay. Dropout is applied both in attention and projection layers for regularization.

This architecture integrates heterogeneous data types, numerical and categorical, into a unified, dense representation. Embeddings allow the model to exploit latent similarities among categorical entities without incurring the cost of high-dimensional sparse vectors. The positional and time-gap embeddings encode absolute sequence position and irregular temporal dynamics, both of which are crucial for interpreting user engagement behavior. The LSTM provides strong modeling of local order-sensitive dependencies, while the causal attention layer captures global relationships, enabling the model to focus selectively on distant but influential events. Finally, the gated projection head acts as a content-aware filter, passing forward only the most relevant transformed features for prediction. Together, these design elements yield a model capable of accurately forecasting engagement while respecting strict causality constraints.

\section{Feature Generations $X$ and Learning Algorithm Implementation}\label{app:Features}
The features $X$ comprise \textbf{contextual} (\(X^C\)), \textbf{geographical} (\(X^{Geo}\)), and \textbf{behavioral} (\(X^{Behave}\)) components. Contextual features include static metadata (e.g., time, device, network), geographical features encode spatial attributes (latitude/longitude, province, city) using low-dimensional embeddings, and behavioral features capture dynamic engagement patterns. We detail feature generation, preprocessing, and the variables associated with each information set in the following subsections.

\subsection{Contextual features $X^C$}
To capture the cyclical nature of time, the hour and minute of the day are encoded using sine and cosine transformations. For hour \( h \) and minute \( m \), these features are defined as:
\[
\text{Hour\_SIN} = \sin\left(2\pi \frac{h}{24}\right), \quad \text{Hour\_COS} = \cos\left(2\pi \frac{h}{24}\right),
\]
\[
\text{Minute\_SIN} = \sin\left(2\pi \frac{m}{60}\right), \quad \text{Minute\_COS} = \cos\left(2\pi \frac{m}{60}\right).
\]
These transformations ensure that \( 23:00 \) and \( 00:00 \) are numerically close, accurately reflecting the cyclical nature of time. For each categorical variable—media package (app) name (\( M \)), device model (\( D \)), brand ID (\( B \)), operator ID (\( O \)), ISP ID (\( I \)), and advertisement ID (\( A \))—embeddings are generated to convert them into numerical representations. While \( M, D, B, O, \) and \( I \) focus on the top 50 categories, \( A \) is restricted to 5 predefined categories:  
\[
f_{C}(c) = 
\begin{cases} 
k, & \text{if } c \in \text{Top 50 categories},\ k \in \{1, \dots, 50\}, \\
0, & \text{otherwise}, \quad \text{for } C \in \{M, D, B, O, I\}, \\
k, & \text{if } c \in \text{5 Categories},\ k \in \{1, \dots, 5\}, \\
0, & \text{otherwise}, \quad \text{for } C = A.
\end{cases}
\]
Here, \( C \) represents each feature set, and \( k \) corresponds to the embedded category index. This encoding reduces complexity by focusing on the most common categories while ensuring advertisement ID (\( A \)) maintains its specific 5-category structure.

\subsection{Geographical features $X^{Geo}$}
Geographical information captures the user’s spatial context at varying levels of granularity. In our setting, these features include precise coordinates, \emph{longitude} (\( \lambda \)) and \emph{latitude} (\( \phi \)), as well as administrative divisions such as \emph{province} (\( P \)) and \emph{city} (\( C \)). The coordinates \((\lambda, \phi)\) provide fine-grained locational information that can be used to compute spatial proximity between users or match impressions to regional market conditions. Province and city identifiers serve as coarser geographical categories that can be linked to aggregated demographic, socioeconomic, or cultural attributes. Because province and city are discrete categorical variables with many unique values, we represent them using learnable embeddings. This transformation assigns each category to a dense, low-dimensional vector in a continuous space, enabling the model to capture similarity patterns between locations without losing category-specific information. Formally:
\[
f_{P}(p) = \mathbf{e}_{p} \in \mathbb{R}^{d_P}, 
\quad
f_{C}(c) = \mathbf{e}_{c} \in \mathbb{R}^{d_C},
\]
where \( \mathbf{e}_{p} \) and \( \mathbf{e}_{c} \) are embedding vectors of dimensions \(d_P\) and \(d_C\), respectively, learned jointly with the predictive model.

Integrating $X^G$ with contextual features $X^C$ can enhance targeting performance in multiple ways. First, location variables may interact with device and network attributes, for example, certain smartphone brands or ISPs may dominate in specific regions, allowing the model to capture joint patterns that are not evident from either source alone. Second, combining spatial context with temporal features can improve the detection of region-specific engagement cycles, such as peak usage hours in different cities. Finally, location information, when paired with contextual metadata such as operator ID or ISP ID, can serve as a proxy for unobserved market segmentation factors, thereby enriching the representation of the user’s latent type \( \omega \) without substantially increasing privacy risks. Models such as gradient-boosted decision trees (e.g., XGBoost) are particularly well-suited for this integration, as they can flexibly capture nonlinearities and high-order interactions between heterogeneous feature types, including continuous coordinates,  and categorical embeddings, without requiring strong parametric assumptions.

\subsection{Behavioral features $X^{Behave}$}
The behavioral features capture user interactions and engagement dynamics at different levels. These features are categorized into three main types: \textit{user-level features}, which track overall user exposure, clicks, and recency of interactions; \textit{ad-level features}, which measure ad performance, frequency, and user engagement with specific ads; and \textit{app-level features}, which focus on app usage, preferences, and their influence on user clicks. Together, these behavioral features provide a comprehensive view of user activity, enabling the model to identify patterns and relationships that drive engagement and response.  

\subsubsection{User Behavioral Features}

\paragraph{Exposure Count (EC):}  
Tracks the cumulative number of ad exposures for user \( i \) up to but not including the current impression \( n \):  
\[
E_{i,n} = n - 1.
\]  
This feature provides a measure of user exposure history, which helps evaluate the frequency of ad interactions over time.

\paragraph{Click History (CH):}  
Represents the cumulative number of clicks by user \( i \) up to but not including the current impression \( n \):  
\[
C_{i,n} = \sum_{k=1}^{n-1} \text{CLICK}_{i,k}.
\]  
By excluding the current click, this feature highlights past user engagement and serves as a strong indicator of historical interaction behavior.

\paragraph{Session Click-Through Rate (SCTR):}  
Measures the user’s CTR within the session, calculated as the ratio of past clicks to past exposures for user \( i \) at impression \( n \):  
\[
S_{i,n} = \frac{C_{i,n}}{E_{i,n}}.
\]  
This feature captures the effectiveness of ads in driving clicks during a session and reflects user engagement intensity.

\paragraph{Time Since Last Exposure (TSE):}  
Represents the time gap between the current and previous ad exposures for user \( i \):  
\[
TSE_{i,n} = \text{TIME}_{i,n} - \text{TIME}_{i,n-1}.
\]  
This feature helps identify the recency of ad interactions, which can influence user responsiveness to repeated exposures.

\paragraph{Time Since Last Click (TCE):}  
Tracks the time elapsed since the last click by user \( i \), up to the current impression \( n \):  
\[
TCE_{i,n} = \text{TIME}_{i,n} - \text{LAST\_CLICK\_TIME}_{i}.
\]  
By measuring click recency, this feature provides insight into how recent engagement influences future click behavior.

\subsubsection{Ad-Level Behavioral Features}

\paragraph{Ad Frequency (F):}  
Represents the cumulative number of times user \( i \) has been exposed to ad \( j \) up to the current impression \( n \):  
\[
F_{i,j,n} = n.
\]  
This feature helps track ad repetition for a specific user, which is critical for understanding the impact of ad frequency on engagement.

\paragraph{Ad Click-Through Rate (Ad CTR):}  
Measures the click-through rate for ad \( j \) by user \( i \), based on clicks up to but not including the current impression \( n \):  
\[
\text{CTR}_{i,j,n} = \frac{\sum_{k=1}^{n-1} \text{CLICK}_{i,j,k}}{n-1}.
\]  
By excluding the current click, this feature captures past user engagement with a specific ad, offering insights into its historical performance.

\paragraph{Overall Ad Click-Through Rate (Overall Ad CTR):}  
Represents the overall click-through rate for ad \( j \) across all users, based on clicks up to but not including the current impression \( n \):  
\[
\text{CTR}_{j}^{\text{Overall}} = \frac{\sum_{i} \sum_{k=1}^{n_{i,j}-1} \text{CLICK}_{i,j,k}}{\sum_{i} (n_{i,j}-1)}.
\]  
This feature benchmarks the performance of an ad across the entire user base, providing a global measure of its effectiveness.

\subsubsection{App-Level Behavioral Features}

\paragraph{Usage Share of App (U):}  
Represents the proportion of exposures for app \( a \) relative to all exposures for user \( i \) up to but not including the current impression \( n \):  
\[
U_{i,a,n} = \frac{\text{Exposures}_{i,a,n}}{\text{Total\_Exposures}_{i,n}}.
\]  
This feature highlights user preferences for specific apps by tracking their exposure distribution across all apps.

\paragraph{Effect of App (E):}  
Measures the contribution of app \( a \) to cumulative user clicks, excluding the current click at impression \( n \):  
\[
E_{i,a,n} = \frac{\text{Clicks}_{i,a,n}}{\text{Total\_Clicks}_{i,n}}.
\]  
By isolating historical click behavior, this feature evaluates the influence of a specific app on user engagement.

\paragraph{Preference for App (P):}  
Represents the user’s preference for app \( a \) within the context of ad \( j \), based on exposures up to but not including the current impression \( n \):  
\[
P_{i,a,j,n} = \frac{\text{Exposures}_{i,a,j,n}}{\text{Total\_Exposures}_{i,j,n}}.
\]  
This feature captures how user interactions with apps vary within specific ad contexts, providing insight into app-specific ad performance.

\paragraph{Influence of App (I):}  
Quantifies the impact of app \( a \) on user clicks within the context of ad \( j \), excluding the current click at impression \( n \):  
\[
I_{i,a,j,n} = \frac{\text{Clicks}_{i,a,j,n}}{\text{Total\_Clicks}_{i,j,n}}.
\]  
This feature measures the effectiveness of an app in driving clicks for a specific ad, offering app-level insights for targeted strategies.

\paragraph{Overall App Usage (\( U^{\text{Overall}} \)):}  
Represents the share of exposures for app \( a \) across all users, based on cumulative exposures up to but not including the current impression \( n \):  
\[
U_{a,n}^{\text{Overall}} = \frac{\text{Exposures}_{a,n}}{\text{Total\_Exposures}_{n}}.
\]  
This feature tracks the overall popularity of an app by monitoring its relative exposure share across the dataset.

\paragraph{Overall App Effect (\( E^{\text{Overall}} \)):}  
Measures the contribution of app \( a \) to overall clicks across all users, excluding the current click at impression \( n \):  
\[
E_{a,n}^{\text{Overall}} = \frac{\text{Clicks}_{a,n}}{\text{Total\_Clicks}_{n}}.
\]  
This feature evaluates the global impact of an app on user engagement, providing a benchmark for app performance.

\section{IPS and Propensity Score Evaluations}
\subsection{Proof of Proposion \ref{prop:ipsVX} (deterministic $\pi^{f_X}$)}\label{app:ips_proof}
Define the deterministic IPS estimator
\[
\hat V(\pi^{f_X})
= \frac{1}{N}\sum_{i=1}^N 
\frac{\mathbf{1}\{A_i=\pi^{f_X}(X_i)\}}{\pi^{\mathcal D}(A_i\mid X_i)} \, v(A_i)\, Y_i,
\]
and treat $\pi^{f_X}$ as fixed w.r.t.\ the evaluation sample (by using a holdout or cross-fitting).
\begin{itemize}
\item \textit{Step 1:}
By i.i.d.\ sampling of $(X_i,A_i,Y_i)$ from the logging DGP $\big(X\sim p_X,\,A\mid X\sim\pi^{\mathcal D}(\cdot\mid X),\,Y\mid X,A\sim \Pr(\cdot\mid X,A)\big)$ and boundedness (Overlap Assumption \ref{ass:overlap} ensures the denominator is nonzero whenever $\mathbf{1}\{A=\pi^f(X)\}=1$; $Y\in\{0,1\}$; $v(\cdot)$ finite), the strong law of large numbers (SLLN) yields
\[
\hat V(\pi^{f_X})\xrightarrow{a.s.}
\mathbb{E}\!\left[\frac{\mathbf{1}\{A=\pi^{f_X}(X)\}}{\pi^{\mathcal D}(A\mid X)}\, v(A)\, Y\right].
\]

\item \textit{Step 2:} First law of iterated expectations condition on $X$ yields:
\[
\mathbb{E}\!\left[\frac{\mathbf{1}\{A=\pi^f(X)\}}{\pi^{\mathcal D}(A\mid X)}\, v(A)\, Y\right]
= \mathbb{E}_X\!\left[
\mathbb{E}\!\left[\frac{\mathbf{1}\{A=\pi^f(X)\}}{\pi^{\mathcal D}(A\mid X)}\, v(A)\, Y \,\middle|\, X\right]
\right].
\]

\item \textit{Step 3:} Second law of iterated expectations condition on $A\mid X$ and expand the sum yields:
\[
\mathbb{E}\!\left[\frac{\mathbf{1}\{A=\pi^{f_X}(X)\}}{\pi^{\mathcal D}(A\mid X)}\, v(A)\, Y \,\middle|\, X\right]
= \sum_{a\in\mathcal A} \pi^{\mathcal D}(a\mid X)\,
\frac{\mathbf{1}\{a=\pi^{f_X}(X)\}}{\pi^{\mathcal D}(a\mid X)} \, v(a)\,
\mathbb{E}[Y\mid X,a].
\]

\item \textit{Step 4:}
The logging propensity cancels given Overlap Assumption \ref{ass:overlap}:
\[
= \sum_{a\in\mathcal A} \mathbf{1}\{a=\pi^{f_X}(X)\}\, v(a)\, \mathbb{E}[Y\mid X,a]
= v(\pi^{f_X}(X))\, \mathbb{E}[Y\mid X, A=\pi^{f_X}(X)].
\]
Therefore,
\[
\mathbb{E}\!\left[\frac{\mathbf{1}\{A=\pi^f(X)\}}{\pi^{\mathcal D}(A\mid X)}\, v(A)\, Y\right]
= \mathbb{E}_X\!\left[ v(\pi^f(X))\, \mathbb{E}[Y\mid X, A=\pi^f(X)] \right].
\]

\item \textit{Step 5:}
By Unconfoundedness Assumption~\ref{ass:unconfoundedness} and the usual consistency ($Y=Y(A)$),
\[
\mathbb{E}[Y\mid X,a] \;=\; \mathbb{E}[Y(a)\mid X].
\]
Hence
\[
\mathbb{E}_X\!\left[ v(\pi^{f_X}(X))\, \mathbb{E}[Y\mid X, A=\pi^{f_X}(X)] \right]
= \mathbb{E}_X\!\left[ v(\pi^{f_X}(X))\, \mathbb{E}[Y(\pi^{f_X}(X))\mid X] \right].
\]

\item \textit{Step 6:} By the definition, $u(a,\omega)=v(a)\Pr(Y=1\mid a,\omega)$ and $q_X(\omega)=\Pr(\omega\mid X)$, so for any $x$ and $a$,
\[
v(a)\,\mathbb{E}[Y(a)\mid X=x]
= v(a)\sum_{\omega}\Pr(Y=1\mid a,\omega)\,q_x(\omega)
= \sum_{\omega} u(a,\omega)\, q_x(\omega).
\]
Plugging $a=\pi^{f_X}(x)$ gives:
\[
\mathbb{E}_X\!\left[ v(\pi^{f_X}(X))\, \mathbb{E}[Y(\pi^{f_X}(X))\mid X] \right]
= \mathbb{E}_X\!\left[ \sum_{\omega} u(\pi^{f_X}(X),\omega)\, q_X(\omega)\right]
= V(\pi^{f_X}).
\]

\item \textit{Step 7:} Combining Steps 1–6,
\[
\mathbb{E}\!\big[\hat V(\pi^{f_X})\big] \;=\; V(\pi^{f_X}),
\qquad
\hat V(\pi^{f_X}) \xrightarrow{a.s.} V(\pi^{f_X}).
\]

\item \textit{Step 8:} If Policy optimality under $X$ Assumption~\ref{ass:optimal} holds, then
\[
V(\pi^{f_X})=\sup_{\pi} V(\pi) \;=:\; V_X.
\]
Therefore,
\[
\mathbb{E}\!\big[\hat V(\pi^{f_X})\big]=V_X,
\qquad
\hat V(\pi^{f_X})\xrightarrow{p} V_X.
\]
\(\square\)

\end{itemize}

\subsection{Propensity Score Estimation and Validation}\label{app:Propensity}
Our IPS implementation requires estimates of the platform’s logging policy probabilities (\emph{propensity scores}) and verification of covariate balance. Appendix \S\ref{app:propensity_estimation} details the estimation procedure, and Appendix \S\ref{app:balance} reports balance diagnostics confirming small post‐weighting differences.

\subsubsection{Propensity Score Estimation}\label{app:propensity_estimation}
\paragraph{Support and Eligibility Filtering.}  
In our setting, not all ads are eligible to appear in every impression due to platform-level targeting constraints such as location, time-of-day, device type, or campaign-specific restrictions. To encode these constraints, we construct an \emph{eligibility matrix} following \cite{Rafieian2021}: 
\[
E \in \{0,1\}^{N \times A},
\]  
where \( e_{i,a} = 1 \) if and only if ad \(a\) is eligible to be shown in impression \(i\). The matrix \(E\) is obtained by intersecting eligibility indicators across multiple categorical dimensions, province, hour-of-day, app category, smartphone brand, network type, and media package name, where each dimension’s support is derived from the set of observed ad–context combinations in the historical data.  

The eligibility matrix \(E\) serves two roles in our estimation procedure. First, the columns \(\{e_{\cdot,a}\}_{a=1}^A\) are included as binary features in the propensity model, allowing the learner to incorporate eligibility information directly into its predictions. Second, after model estimation, we enforce the \emph{support condition} by setting \(\hat{\pi}^{\mathcal{D}}_{i,a} = 0\) whenever \(e_{i,a} = 0\), followed by renormalization across eligible ads. This guarantees that  
\[
\hat{\pi}^{\mathcal{D}}(a \mid x_i) > 0 \quad \text{only if} \quad e_{i,a} = 1,
\]  
while still permitting the model to flexibly learn variation in serving probabilities within the support.

\paragraph{Outcome and Covariates.}  
The propensity score model predicts, for each impression \(i \in \{1,\dots,N\}\), which ad from the candidate set \(\mathcal{A} = \{a^{(1)}, a^{(2)}, \dots, a^{(K)}\}\) was displayed. The outcome variable is the multi-class indicator \(\mathbf{1}\{A_i = a^{(j)}\}\), where \(A_i\) denotes the ad shown in impression \(i\).  

The covariate vector \(x_i\) is constructed to capture all observed determinants of ad allocation and includes the following components:  

\begin{enumerate}
    \item \textbf{Eligibility indicators:} For each ad \(a\), we include the binary feature \(e_{i,a} \in \{0,1\}\) from the eligibility matrix \(E\) described above. This ensures that the model incorporates ad-specific availability and targeting rules directly into its predictions.  
    \item \textbf{Targeting variables:} Province, app category, hour-of-day, smartphone brand, network connection type, and mobile service provider, each factorized into integer-encoded indices to capture categorical variation.
    \item \textbf{Fine-grained location:} Latitude and longitude, enabling the model to detect geographical heterogeneity beyond province-level effects.
    \item \textbf{Temporal controls:} Time-of-day features measured at the minute level, absorbing short-run fluctuations in bids or quality scores due to diurnal cycles or transient platform shocks.
    \item \textbf{Device and context characteristics:} Device identifiers, fraud detection codes, and media package identifiers, capturing residual variation in serving probabilities not explained by higher-level targeting variables.
\end{enumerate}
All categorical covariates are transformed into numeric indices using factorization, preserving a consistent mapping across impressions. The eligibility indicators \(e_{i,a}\) for all \(a \in \mathcal{A}\) are appended to the base covariate set, so the final feature matrix combines both contextual features and ad-specific availability indicators.

\paragraph{Estimation Procedure.}  
The platform’s quasi‐proportional allocation rule implies that the probability of displaying ad \(a \in \mathcal{A}\) in impression \(i\) can be expressed as  
\[
\pi^{\mathcal{D}}(a \mid x_i) \;=\; 
\frac{b_a(x_i) \, q_a(x_i)}
{\sum_{a' \in \mathcal{A}_i} b_{a'}(x_i) \, q_{a'}(x_i)},
\]  
where \(b_a(x_i)\) and \(q_a(x_i)\) denote, respectively, the bid and quality score functions conditional on the observed covariates \(x_i\), and \(\mathcal{A}_i \subseteq \mathcal{A}\) is the set of ads eligible in context \(x_i\).  

In practice, these bid and quality components are not fully observed, and the realized allocation process may deviate from the theoretical form due to unobserved adjustments, platform‐level interventions, and complex interactions among covariates. We therefore estimate the realized propensities \(\pi^{\mathcal{D}}\) directly from the historical allocation data using a flexible, nonparametric method. Specifically, we adopt a {binary one‐vs‐all} approach: for each \(a^{(j)} \in \mathcal{A}\), we estimate the conditional probability  
\[
\pi^{\mathcal{D}}_{j}(x_i) \;=\; \Pr\!\big(A_i = a^{(j)} \,\big|\, x_i\big),
\]  
By estimating a separate binary classifier for each ad \(a^{(j)}\), we capture ad‐specific allocation patterns without imposing a single functional form across all ads—an important flexibility given heterogeneous eligibility rules, competition, and campaign objectives. This one‐vs‐all formulation also mitigates the severe class imbalance of large‐scale ad data: rare ads, which would exert little influence in a multiclass model’s joint decision boundary, are given dedicated models in which their appearances are positive events. Moreover, it avoids the restrictive multiclass assumption that the same covariate effects and interactions govern all alternatives equally, an implausible condition in our setting.

For each ad‐specific classifier, the training set includes all \(N\) impressions, even those in which the ad was ineligible (\(e_{i,a^{(j)}} = 0\)), allowing the model to learn to assign zero probability in contexts where the ad cannot be served. Each classifier is implemented as an XGBoost with a logistic loss for binary outcomes. To balance predictive accuracy and computational efficiency in a high‐dimensional feature space, we adopt hyperparameters for the learning algorithm XGBosst motivated by \cite{Rafieian2021}.

To ensure that estimated propensities are free from in‐sample overfitting, we adopt a \(K\)-fold cross‐fitting procedure with \(K = 5\). Let the set of impressions be indexed by \(\{1, \dots, N\}\), and let \(\mathcal{I}_1, \dots, \mathcal{I}_K\) denote a partition of this set into \(K\) disjoint folds of (approximately) equal size. For each fold \(\mathcal{I}_k\), the ad‐specific binary classifiers are trained on the complement set \(\bigcup_{m \neq k} \mathcal{I}_m\), and predictions \(\hat{\pi}^{\mathcal{D}}_{i,a}\) are generated only for impressions \(i \in \mathcal{I}_k\). By construction, every propensity score is obtained from a model that did not observe the corresponding impression during training. Concatenating these out‐of‐fold predictions across all \(k = 1, \dots, K\) yields a complete \(N \times A\) matrix of out‐of‐sample estimates. This procedure preserves the predictive structure learned by the models while eliminating the upward bias in inverse propensity estimators that arises from reusing training predictions.

\paragraph{Post Processing.} Following propensity estimation, we align the predictions with the platform’s targeting constraints by imposing the \emph{support condition}. Given \(E \in \{0,1\}^{N \times A}\) be the eligibility matrix, where \(e_{i,a} = 1\) iff ad \(a\) is eligible in impression \(i\). We set  
\[
\hat{\pi}^{\mathcal{D}}_{i,a} = 0 \quad \text{for all} \quad a \notin \mathcal{A}_i,
\]
where \(\mathcal{A}_i = \{a \in \mathcal{A} : e_{i,a} = 1\}\) is the eligible set for impression \(i\). The remaining probabilities are renormalized:  
\[
\hat{\pi}^{\mathcal{D}}_{i,a} \; \leftarrow \; 
\frac{\hat{\pi}^{\mathcal{D}}_{i,a}}
{\sum_{a' \in \mathcal{A}_i} \hat{\pi}^{\mathcal{D}}_{i,a'}} 
\quad \forall a \in \mathcal{A}_i,
\]
so that \(\sum_{a \in \mathcal{A}_i} \hat{\pi}^{\mathcal{D}}_{i,a} = 1\) and  
\[
\hat{\pi}^{\mathcal{D}}(a \mid x_i) > 0 \iff e_{i,a} = 1.
\]

This adjustment preserves the relative allocation probabilities learned by the model among eligible ads while ensuring no probability mass is assigned to ineligible alternatives.

\paragraph{Results from Propensity Score Estimation.}  
The estimated propensities capture systematic variation in ad allocation without overfitting to deterministic rules. The macro‐averaged AUC is \(0.6335\) and the micro‐averaged AUC is \(0.6276\), indicating moderate discriminatory power above random guessing but far from perfect classification. This aligns with expectations: if the platform’s allocation contains substantial quasi‐randomization (subject to eligibility and proportional bidding), a very high AUC would be undesirable, as it would imply near‐deterministic serving and reduced overlap. Likewise, the macro and micro log‐loss values (\(0.4862\) and \(0.4388\)) are low enough to indicate well‐calibrated predictions but not so low as to suggest overconfidence.

\begin{figure}[htp!]
    \centering
    \includegraphics[width=0.8\linewidth]{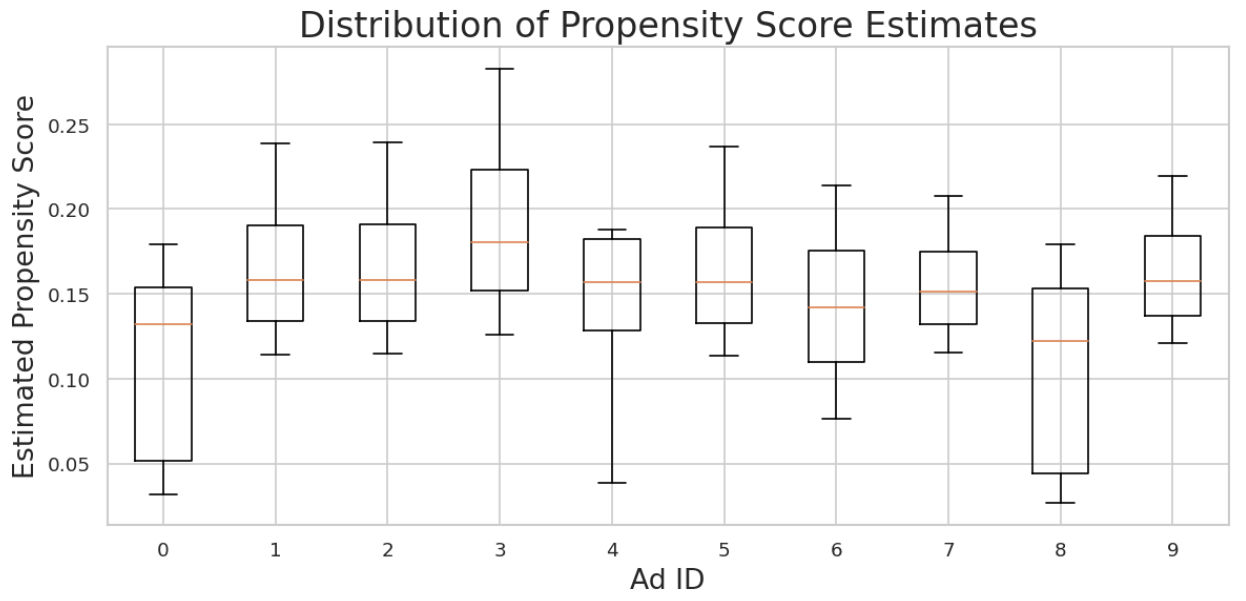}
    \caption{Distribution of estimated propensity scores across ads. Dispersion and non‐degeneracy indicate adequate common support.}
    \label{fig:boxplot_propensity}
\end{figure}

Figure~\ref{fig:boxplot_propensity} plots the distribution of estimated propensities for each ad. For most ads, the median serving probability lies between \(0.12\) and \(0.18\), with interquartile ranges spanning several percentage points. This dispersion indicates that the model assigns a range of probabilities across impressions rather than concentrating mass at extreme values, which is essential for maintaining overlap. Crucially, no ad exhibits degenerate propensities concentrated near zero (indicating near‐never assignment) or near one (indicating near‐deterministic assignment). Even the rarest ads display a nontrivial upper tail, implying that all ads have a positive probability of being shown in multiple contexts. To further assess whether reweighting achieves covariate balance across treatment arms, we next compute the standardized bias (SB) before and after applying the estimated inverse propensity weights in \S\ref{app:balance}.

\subsubsection{Covariate Balance Diagnostics}\label{app:balance}
After estimating the logging policy propensities \(\hat{\pi}^{\mathcal{D}}_{i,a}\) and enforcing support, we evaluate whether inverse‐propensity weighting balances the distribution of each covariate across treatment arms using the \emph{standardized bias} (SB) metric. The goal is to check whether, after weighting, the covariate distribution for impressions assigned to any given ad matches that of the eligible population, thereby providing evidence consistent with the unconfoundedness assumption under overlap. Let \(X\) denote a generic covariate, \(X_i\) its value for impression \(i\), and \(A_i\) the ad shown. For ad \(a \in \mathcal{A}\), the \emph{unweighted} mean of \(X\) in impressions where \(A_i = a\) is
\[
\bar{X}_a \;=\; \frac{\sum_{i=1}^N \mathbf{1}(A_i = a) \, X_i}{\sum_{i=1}^N \mathbf{1}(A_i = a)},
\]
while the mean in the \emph{eligible} population is
\[
\bar{X} \;=\; \frac{1}{N} \sum_{i=1}^N X_i.
\]
Let \(\sigma_X\) be the standard deviation of \(X\) in the eligible population. The \emph{unweighted} standardized bias for \(X\) when assigned to \(a\) is then
\[
SB_a(X) \;=\; \frac{\left|\,\bar{X}_a - \bar{X}\,\right|}{\sigma_X},
\]
and the worst‐case imbalance for \(X\) across all ads is
\[
SB(X) \;=\; \max_{a \in \mathcal{A}} SB_a(X).
\]
Intuitively, \(SB_a(X)\) measures the relative shift (in standard deviation units) between the covariate mean for impressions shown ad \(a\) and the overall eligible population mean; \(SB(X)\) takes the largest such shift across ads. Following \citet{mccaffrey2013tutorial}, we adopt the benchmark \(|SB(X)| < 0.20\) as the criterion for acceptable balance. Values above this threshold indicate potentially meaningful imbalance in \(X\) between treatment arms. After weighting by the inverse of the estimated propensities, the weighted mean of \(X\) for impressions shown ad \(a\) is
\[
\bar{X}^{\hat{\pi}}_{a} \;=\; \frac{\sum_{i=1}^N \mathbf{1}(A_i = a) \,\frac{X_i}{\hat{\pi}^{\mathcal{D}}_{i,a}}}{\sum_{i=1}^N \mathbf{1}(A_i = a) \,\frac{1}{\hat{\pi}^{\mathcal{D}}_{i,a}}}.
\]
Let \(\bar{X}^{\mathrm{elig}(a)} = \frac{1}{|\mathcal{S}_a|}\sum_{i \in \mathcal{S}_a} X_i\) and \(\sigma_X^{\mathrm{elig}(a)}\) denote, respectively, the mean and standard deviation of \(X\) in the ad‐specific eligible set \(\mathcal{S}_a\) (i.e., rows with support for ad \(a\)). The \emph{post‐weighting} standardized bias for \(X\) when assigned to \(a\) is
\[
SB_a^{\hat{\pi}}(X) \;=\; \frac{\left|\,\bar{X}^{\hat{\pi}}_{a} - \bar{X}^{\mathrm{elig}(a)}\,\right|}{\sigma_X^{\mathrm{elig}(a)}},
\]
and the worst‐case post‐weighting imbalance is
\[
SB^{\hat{\pi}}(X) \;=\; \max_{a \in \mathcal{A}} SB_a^{\hat{\pi}}(X).
\]

\begin{figure}[htp!]
    \centering
    \includegraphics[width=0.60\linewidth]{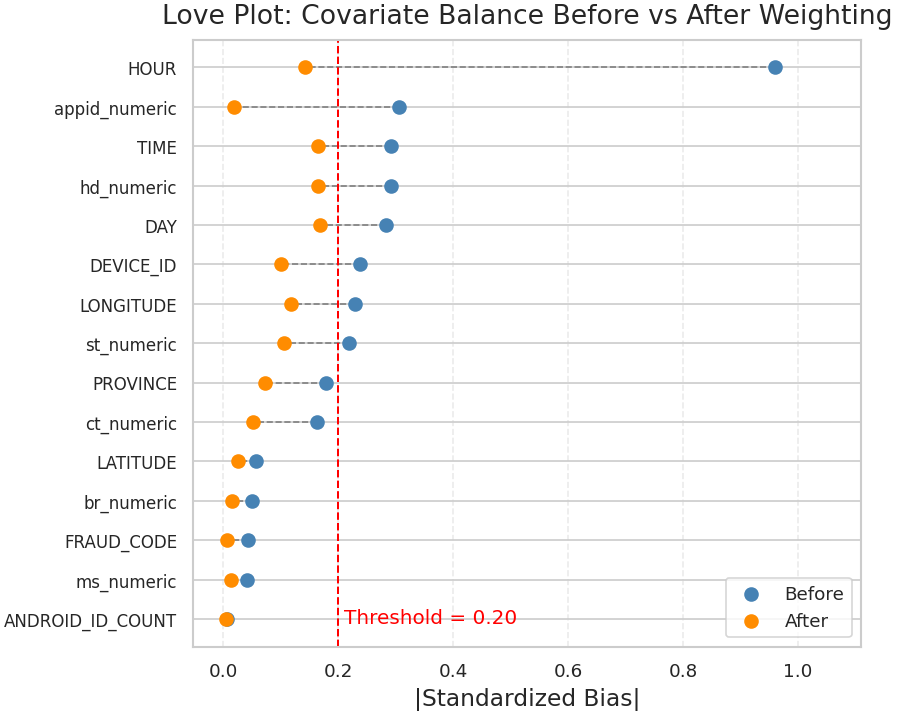}
    \caption{Standardized bias for each covariate before and after inverse‐propensity weighting. The red dashed line marks the \(0.20\) balance threshold from \citet{mccaffrey2013tutorial}.}
    \label{fig:love_plot}
\end{figure}

If \(SB^{\hat{\pi}}(X)\) satisfies the balance threshold for every \(X\), then the inverse‐propensity weighting has successfully aligned the covariate distributions across all treatment arms (conditional on the enforced support). Comparing \(SB(X)\) (pre‐weighting) and \(SB^{\hat{\pi}}(X)\) (post‐weighting) quantifies the extent to which our weighting procedure mitigates initial selection bias. Figure~\ref{fig:love_plot} reports the worst‐case standardized bias \(SB(X)\) and its post‐weighting counterpart \(SB^{\hat{\pi}}(X)\) for each covariate, as defined. 

Before weighting (blue), several covariates, most notably {HOUR}, exhibit large imbalances, with \(SB(X) > 0.20\), indicating substantial divergence between the covariate distribution for impressions assigned to certain ads and that of the eligible population. After inverse‐propensity weighting (orange), all covariates satisfy the balance criterion, i.e., \(SB^{\hat{\pi}}(X) < 0.20\). This demonstrates that reweighting has effectively aligned the covariate distributions across treatment arms, substantially mitigating the selection bias present in the unweighted data. The improvement is consistent across both high‐variance contextual features (e.g., {PROVINCE}, {LONGITUDE}) and stable identifiers (e.g., {DEVICE\_ID}), providing empirical support for the overlap assumption in our setting.  

\section{Additional Empirical Results}
\subsection{XGBoost vs. LSTM in Predictive Performance}\label{app:XGboostLSTM_Prediction}
To benchmark our models against approaches used in the literature, we implement an XGBoost model following the specifications in \citet{Rafieian2021}. This allows us to directly compare the predictive performance of our LSTM model with a widely‐used tree‐based method in ad‐targeting research.  

For both behavioral ($X^B$) and combined ($X^{GB}$) feature sets, we train the XGBoost model using the same training/test splits, hyperparameter tuning protocol, and feature preprocessing pipeline applied to our LSTM models. Hyperparameters are selected via grid search to maximize AUC on the validation set, and class weights are adjusted to account for click sparsity. Table~\ref{tab:auc_rig_xgb_lstm} reports performance metrics for the two methods.  

\begin{table}[htp!]
\centering
\begin{tabular}{lccc}
\toprule
\textbf{Model} & \textbf{Log Loss} & \textbf{AUC} & \textbf{Relative Information Gain} \\
\midrule
XGBoost ($X^B$)    & 0.0552 & 0.8235 & 37.64\% \\
XGBoost ($X^{GB}$) & 0.0549 & 0.8333 & 37.98\% \\
LSTM ($X^B$)    & 0.0140 & 0.8088 & 84.18\% \\
LSTM ($X^{GB}$) & 0.0138 & 0.8115 & 84.41\% \\
\bottomrule
\end{tabular}
\caption{Comparison of model performance for behavioral ($X^B$) and combined ($X^{GB}$) features using XGBoost and LSTM model. RIG computed as baseline CTR of test set is $p=0.017592$.}
\label{tab:auc_rig_xgb_lstm}
\end{table}
The results reveal three key patterns. First, consistent with our earlier findings, behavioral features provide the dominant source of predictive power. Second, adding geographical information to behavioral features yields only marginal improvements, confirming that behavioral data capture most of the variation in click propensity. Third, the LSTM model substantially outperforms XGBoost in log loss and RIG for the same feature set, despite almost similar AUC values. This indicates that the LSTM’s ability to model temporal dependencies in behavioral sequences translates into better probabilistic calibration, even if its rank‐ordering performance (AUC) is comparable to the tree‐based benchmark.  

It is important to emphasize that high predictive performance does not necessarily translate into higher decision value in targeting. A model may accurately rank or calibrate predicted click probabilities but still yield limited improvements in actual campaign outcomes under realistic budget and exposure constraints. In the next section, we therefore evaluate these models in terms of their realized \emph{decision value} as measured by the targeting value function.

\subsection{XGBoost vs. LSTM in Value of Information Performance}\label{app:XGboostLSTM_Value}
To complement the predictive performance comparison in Appendix \S\ref{app:XGboostLSTM_Prediction}, we benchmark the \emph{decision value} delivered by LSTM and XGBoost models. Decision value is measured using the IPS‐based targeting value function, which quantifies the incremental CTR lift relative to the logging‐policy baseline under realistic targeting constraints. This analysis addresses the fact that models with similar predictive accuracy can differ in their ability to generate value when deployed in practice.

For both behavioral ($X^B$) and combined ($X^{GB}$) feature sets, we use the same training/test partitions and preprocessing as in the predictive exercise, ensuring that differences in decision value are attributable to the learning algorithm rather than data handling. IPS estimates are computed following the same procedure described in \S\ref{ssec:ips_estimation}, and standard errors are clustered at the user level. Table~\ref{tab:policy_value_ips2} reports the estimated policy values, confidence intervals, $t$‐statistics, standard errors, percentage lifts over the baseline CTR of $p=0.017592$, and effective sample sizes.

\begin{table}[htbp]
\centering
\caption{Estimated Policy Value Using IPS}
\label{tab:policy_value_ips2}
\begin{tabular}{lcccccc}
\toprule
Policy & IPS Estimate & 95\% CI & t-stat & SE & Lift (\%) & ESS \\
\midrule
$\hat{V}_{X^{GB}}$ LSTM        & 0.024883$^{***}$ & [0.024366, 0.025401] & 27.611 & 0.000264 & 41.45\% & 760316 \\
$\hat{V}_{X^{GB}}$ XGBoost     & 0.023531$^{***}$ & [0.023056, 0.024006] & 24.521 & 0.000242 & 33.76\% & 803214 \\
$\hat{V}_{X^B}$ LSTM           & 0.022918$^{***}$ & [0.022387, 0.023448] & 19.675 & 0.000271 & 30.28\% & 555871 \\
$\hat{V}_{X^B}$ XGBoost        & 0.022738$^{***}$ & [0.022261, 0.023214] & 21.183 & 0.000243 & 29.25\% & 735365 \\
\bottomrule
\end{tabular}
\\[1.ex]
\begin{minipage}{0.9\textwidth}
\footnotesize
\textit{Notes:} All estimates are computed using IPS. The baseline CTR under the logging policy is 0.017592. Lift is computed relative to this baseline. SE denotes the standard error of the estimate. Effective sample size (ESS) measures the number of equally weighted observations that would yield equivalent precision.  
Number of observations: 3,162,376. Cluster‐robust standard errors are computed using 141,595 user clusters.  
$^{*}p<0.05$, $^{**}p<0.01$, $^{***}p<0.001$
\end{minipage}
\end{table}

The results yield two main insights. First, the LSTM model consistently achieves higher IPS estimates than XGBoost for the same feature set. This gap is particularly pronounced for the combined feature set, where LSTM delivers a 41.45\% lift versus 33.76\% for XGBoost. These differences align with the log‐loss and RIG advantages observed in the predictive exercise, suggesting that the LSTM’s temporal‐sequence modeling translates into superior decision value in deployment. Notably, while XGBoost achieves comparable, and in the case of AUC, slightly higher rank‐ordering performance in the predictive task, these apparent advantages do not translate into higher realized value. This underscores that strong predictive metrics do not guarantee superior targeting outcomes. Second, the divergence between predictive and decision‐value rankings highlights the importance of evaluating targeting algorithms using realized performance metrics, not solely predictive scores. While AUC and calibration remain useful for model assessment, the ultimate measure of a targeting model’s reward is the incremental value it generates in live allocation settings. 

\subsection{Statistical Test of Complementarity vs.\ Substitutability}\label{app:ComSubTests}

Here we present the detailed statistical results underlying Figure~\ref{fig:comsub_depth}. The figure shows mean incremental policy lifts across impression depth, while Table~\ref{tab:comp_sub_test_full} reports the corresponding formal test of whether geographical data complements or substitutes behavioral data at different stages of exposure. The test logic follows a difference-in-differences (DoD) design. Specifically, we define
\[
\hat{\Delta} = (V^{GB}-V^B) - (V^G - V^{\varnothing}),
\]
where $V^{GB}$ is the value of the combined geographical + behavioral policy, $V^B$ is behavior only, $V^G$ is geographical + context, and $V^{\varnothing}$ is context only. Intuitively, $\hat{\Delta}$ compares the incremental contribution of geographical data when behavioral information is present versus when it is absent. A positive $\hat{\Delta}$ indicates that geographical acts as a \emph{complement} to behavior (its marginal value is higher once behavior is available), while a negative $\hat{\Delta}$ implies \emph{substitutability} (its marginal value declines once behavior is available).  

Inference is based on cluster-robust standard errors at the user level. This adjustment ensures that repeated impressions from the same individual do not bias the test statistics. For each impression-depth bin, the table reports the point estimate $\hat{\Delta}$, its $95\%$ confidence interval, the $t$-statistic, a two-sided $p$-value, and one-sided bootstrap probabilities $p(\Delta > 0)$ and $p(\Delta < 0)$. The final two columns summarize the decision rule (Complement, Substitute, or Inconclusive) and the corresponding significance tier.  

The results confirm the visual evidence from Figure~\ref{fig:comsub_depth}: during early exposures (up to roughly 20--25 impressions), $\hat{\Delta}$ is positive and statistically significant, indicating that geographical data provide complementary value when behavioral histories are sparse. Beyond 50 impressions, $\hat{\Delta}$ becomes negative and significant, showing that geographical data turns into a substitute once behavioral data are sufficiently rich. Mid-range bins (around 30--40 impressions) produce inconclusive or transitional estimates. Together, the figure and table establish a dynamic pattern: geographical data is initially complementary, but ultimately substitutable as behavioral histories accumulate.

\begin{table}[htbp]
\centering
\begin{threeparttable}
\caption{Complement vs.\ Substitute by Impression Depth (Clustered DoD Test)}
\label{tab:comp_sub_test_full}
\begin{tabular}{lcccccccc}
\toprule
Last-Impr & $\hat{\Delta}$ & 95\% CI & $t$-stat & $p$ (two) & $p(\Delta>0)$ & $p(\Delta<0)$ & Decision & Signif. \\
\midrule
3   &  0.0006 & [--0.0017, 0.0029] &  0.50 & 0.616 & 0.308 & 0.692 & Inconclusive & none \\
6   &  0.0033 & [0.0008, 0.0057]   &  2.63 & 0.009 & 0.004 & 0.996 & Complement   & strong \\
10  &  0.0023 & [--0.0001, 0.0046] &  1.86 & 0.063 & 0.032 & 0.968 & Complement   & mod \\
16  &  0.0024 & [0.0003, 0.0045]   &  2.23 & 0.026 & 0.013 & 0.987 & Complement   & mod \\
24  &  0.0033 & [0.0014, 0.0052]   &  3.37 & 0.001 & 0.000 & 1.000 & Complement   & strong \\
35  & --0.0002 & [--0.0021, 0.0017] & --0.20 & 0.839 & 0.581 & 0.419 & Inconclusive & none \\
50  & --0.0023 & [--0.0039, --0.0006] & --2.69 & 0.007 & 0.996 & 0.004 & Substitute & strong \\
71  & --0.0019 & [--0.0034, --0.0004] & --2.49 & 0.013 & 0.994 & 0.006 & Substitute & strong \\
102 & --0.0010 & [--0.0024, 0.0003] & --1.54 & 0.123 & 0.938 & 0.062 & Substitute & weak \\
150 & --0.0010 & [--0.0024, 0.0005] & --1.31 & 0.189 & 0.905 & 0.095 & Substitute & weak \\
\bottomrule
\end{tabular}
\begin{tablenotes}[para,flushleft]
\footnotesize
\textit{Notes:} $\hat{\Delta} = (V^{GB}-V^B) - (V^G - V^{\varnothing})$ reports the difference-in-differences estimate of complementarity versus substitutability.  
$t$-statistics and confidence intervals are based on cluster-robust standard errors at the user level. $p(\Delta>0)$ and $p(\Delta<0)$ are one-sided bootstrap probabilities.  
\end{tablenotes}
\end{threeparttable}
\end{table}

\end{appendices}

\end{document}